\documentclass[times]{nmeauth}

\usepackage{amssymb}
\usepackage{amsmath}
\usepackage{cancel}
\usepackage{amsfonts}
\usepackage{bm}

\def \Rdof {{\mathrm{dof}}}

\def \ds {\,\Rd s}

\def \dsb {\,\Rd \bar{s}}
\def \sb {\bar{s}}

\def \Ract {{\mathrm{act}}}

\def \B        #1{{\mbox{$\mathbf #1$}}}      % bold

\def \dO    {\,\Rd\OM}

\def \ds    {\,\Rd s}

\def \pt  {{\partial}}
\newcommand{\fracpt}[2]{\frac{\pt #1}{\pt #2}}
\newcommand{\fracdif}[2]{\frac{\mathrm{d} #1}{\mathrm{d} #2}}

\def \Rd  {{\mathrm{d}}}

\def \Rm  {{\mathrm{m}}}

\def \Rs  {{\mathrm{s}}}

\def \Ru  {{\mathrm{u}}}

\def \RT  {{\mathrm{T}}}

\def \Rint    {{\mathrm{int}}}
\def \Rext    {{\mathrm{ext}}}

\def \Rele    {{\mathrm{ele}}}

\def \Rel     {{\mathrm{el}}}

\def \BB      {{\mathbf B}}
\def \BC      {{\mathbf C}}
\def \BD      {{\mathbf D}}
\def \BE      {{\mathbf E}}
\def \BF      {{\mathbf F}}
\def \BG      {{\mathbf G}}
\def \BH      {{\mathbf H}}
\def \BI      {{\mathbf I}}

\def \BK      {{\mathbf K}}

\def \BN      {{\mathbf N}}

\def \BR      {{\mathbf R}}
\def \BS      {{\mathbf S}}

\def \Bd      {{\mathbf d}}

\def \Bf      {{\mathbf f}}
\def \Bg      {{\mathbf g}}

\def \Bk      {{\mathbf k}}

\def \Bs      {{\mathbf s}}

\def \al      {{\alpha}}

\def \de      {{\delta}}

\def \la      {{\lambda}}

\def \DE      {{\Delta}}

\def \PI      {{\Pi}}

\def \OM      {{\Omega}}
\def \Bal   {{\boldsymbol \alpha}}

\def \Bla   {{\boldsymbol \lambda}}

\def\volumeyear{2020}

\begin{document}

\runningheads{R.~Sachse, F.~Geiger and M.~Bischoff}{Constrained motion design with distinct actuators and motion stabilization}

\title{Constrained motion design with distinct actuators and motion stabilization}

\author{Renate~Sachse\affil{1}$^,$\corrauth, Florian~Geiger\affil{1} and Manfred~Bischoff\affil{1}}

\address{\centering
\affilnum{1}\ Institute for Structural Mechanics, University of Stuttgart, \\ Pfaffenwaldring 7, D-70550 Stuttgart, Germany}

\corraddr{Renate~Sachse, Institute for Structural Mechanics, University of Stuttgart, Pfaffenwaldring 7, D-70550 Stuttgart, Germany, E-mail: sachse@ibb.uni-stuttgart.de}

%%%%%%%%%%%%%%%%%%%%%%%%%%%%%%%%%%%%%%%%%%%%%%%%%%%%%%%%%%%%%
% Abstract

\begin{abstract}
The design of adaptive structures is one method to improve sustainability of buildings. Adaptive structures are able to adapt to different loading and environmental conditions or to changing requirements by either small or large shape changes. In the latter case, also the mechanics and properties of the deformation process play a role for the structure's energy efficiency. The method of variational motion design, previously developed in the group of the authors, allows to identify deformation paths between two given geometrical configurations that are optimal with respect to a defined quality function. In a preliminary, academic setting this method assumes that every single degree of freedom is accessible to arbitrary external actuation forces that realize the optimized motion. These (nodal) forces can be recovered a posteriori. The present contribution deals with an extension of the method of motion design by the constraint that the motion is to be realized by a predefined set of actuation forces. These can be either external forces or prescribed length chances of discrete, internal actuator elements. As an additional constraint, static stability of each intermediate configuration during the motion is taken into account. It can be accomplished by enforcing a positive determinant of the stiffness matrix.
\end{abstract}


\maketitle

%%%%%%%%%%%%%%%%%%%%%%%%%%%%%%%%%%%%%%%%%%%%%%%%%%%%%%%%%%%%%
% Introduction

\section{Introduction}
\label{sec:intro}

Construction industry is responsible for a substantial part of global energy consumption and requirement of material resources. Therefore, the improvement of energy efficiency and sustainability in this field represents a major challenge for architects and engineers, who are determined to design extremely efficient structures. Adaptive structures are one technology to improve structural efficiency by adapting to changing circumstances.

Two types of adaptive structures can be distinguished. The first type, sometimes denoted as smart structure, adapts its internal forces or deformation to varying loads. With the use of sensors and actuators, a light weight design can be accomplished while maintaining structural performance. In doing so, forces and deformations counteracted by actuation~\cite{housner_structural_1997,sobek_adaptive_2001,spencer_b._f._state_2003,korkmaz_review_2011}. Various types of actuation are used, e.g., piezoelectric elements~\cite{irschik_review_2002} or discrete actuator elements in truss systems, which are able to adjust their length. The question of where to place these actuators as efficiently as possible is answered by methods for actuator placement~\cite{abdullah_placement_2001,gupta_optimization_2010,masching_parameter_2016,reksowardojo_actuator_2018}. In this type of adaptive structures, actuation typically induces only small displacements, such that linear structural analyses suffices.

The second type of adaptive structures adapts to environmental changes or to varying service demands through major shape changes. Here, the individual geometric configurations vary significantly. Examples are retractable roofs of stadiums (e.g. the Commerzbank-Arena in Frankfurt, Germany~\cite{goppert_spoked_2007}) or closing and opening of facade elements (e.g. the biomimetic facade element Flectofin~\cite{lienhard_flectofin:_2011}), which contribute to energy efficiency of a building. But also in aviation, shape changes by morphing wings to improve efficiency of airplanes are investigated ~\cite{maute_integrated_2006,liebe_shape-adaptive_2006,vasista_realization_2012,weisshaar_morphing_2013,ajaj_morphing_2016}. The design of flexible or morphing structures that allow for such large geometry changes poses a major challenge. Classically, mechanisms such as folding or sliding with the help of joints are used. For discrete structural typologies, technologies that use length-varying actuator elements have been developed, for instance for truss structures \cite{balaguer_development_2008,senatore_synthesis_2019} and tensegrity structures~\cite{graells_rovira_control_2009,wijdeven_shape_2005,sychterz_deployment_2018}. In the case of continuous structures, such as shells, optimization procedures can be used to increase compliance~\cite{sigmund_design_1997,kota_design_2001,lu_effective_2005,campanile_modal_2008,pagitz_shape-changing_2013,masching_parameter_2016}, thus facilitating continuous bending deformation instead of discrete hinges.

However, not only the geometry of the morphing structure is important for its efficiency. The deformation process has to meet certain requirements and the actuators that control it consume energy. The task of efficient actuation is tackled in control theory, especially in optimal control and robotics. However, pertinent methods mostly focus on rigid systems with discrete kinematics or flexible multibody systems. This is different in the field of continuum robots and hyper-redundant manipulation~\cite{rus_design_2015}, but the continuity and thus a large number of degrees of freedom still poses a challenge. Nevertheless, mechanics and structural analysis have already been combined with optimal control procedures~\cite{ibrahimbegovic_optimal_2004,masic_path_2005,veuve_adaptive_2017}.

The topic of the present study is the design of deformation processes between an initial geometry and a prescribed target geometry that are optimal with respect to a chosen objective function. In a previous work by the authors~\cite{sachse_motion_2019}, the method of \emph{variational motion design} has been presented. It is based on a variational formulation and was developed using an exemplary objective function, namely the internal energy integrated along the deformation path. In robotics, the quantity ``cost of transport'' allows to quantitatively compare different types of transport between two locations, e.g. flying or floating. This quantity can be reinterpreted for deformable structures in motion design to a ``cost of deformation'', which represents a measure for the required energy to deform the structure into the desired target geometry. The method was verified and validated by means of some numerical benchmarks and experiments. A focus was laid on studies of rigid body motions, kinematic structures and inextensional deformations of shells. A brief description of the method is given in Section~\ref{sec:motion_design}.

The method of motion design provides an optimized deformation trajectory that minimizes a certain cost function (e.g. the integrated internal energy). A major hypothesis in this context is that all degree of freedom are accessible to actuation by prescribed forces. These can be computed a posteriori from equilibrium, but of course technical realization of such an omnipotent actuation is rather academic. It is one of the objectives of the present study to remove this limitation. The restriction to a given set of potential actuator forces represents a constraint within the variational framework to compute the resulting motion. 

The constraint that the motion is to be enabled with a prescribed set of actuation forces is enforced by an extended functional with constraints using the Lagrange multiplier method, as explained in Section~\ref{sec:con_load_cases}. Also length changes of discrete actuator elements can be considered as actuation modes. For this purpose, Section~\ref{sec:con_actors} first introduces an actuator element formulation that is suitable for implementation into the framework of motion design. Furthermore, also other constraints may be introduced in this manner. In particular, static stability of every intermediate state during deformation can be guaranteed by enforcing a positive determinant of the stiffness matrix. This overall procedure as well as a combination of both types of constraints (actuation and stability) are presented in Section~\ref{sec:con_detK}.

%%%%%%%%%%%%%%%%%%%%%%%%%%%%%%%%%%%%%%%%%%%%%%%%%%%%%%%%%%%%%

\section{Motion design of structures}
\label{sec:motion_design}

The bases of variational motion design is the functional $J$. As objective function it represents the property that is to be assigned to the motion. The procedure is described in detail
in~Sachse~and~Bischoff~\cite{sachse_motion_2019} and only the most important basics are repeated herein. Geometrically non-linear, quasi-static structural behavior is assumed. Following \cite{sachse_motion_2019}, the internal energy, integrated along the motion path $s$ is considered. With the Green-Lagrange strain $\BE$ and a linear elastic St.~Venant-Kirchhoff material law this leads to
\begin{align}
J &= \int_s \PI_\Rint \ds = \int_s \int_\OM \frac 12 \BE^\RT \BC \BE \dO \ds = \min .
\label{eq:functional}
\end{align}
Alternative objective functions can be used.

\begin{figure}[b]
\centering
%\includesvg{fig_motion_design}
%\includegraphics[width=1.0\textwidth]{Figure_1}
\hspace*{-4mm}
\includegraphics[width=15.0cm]{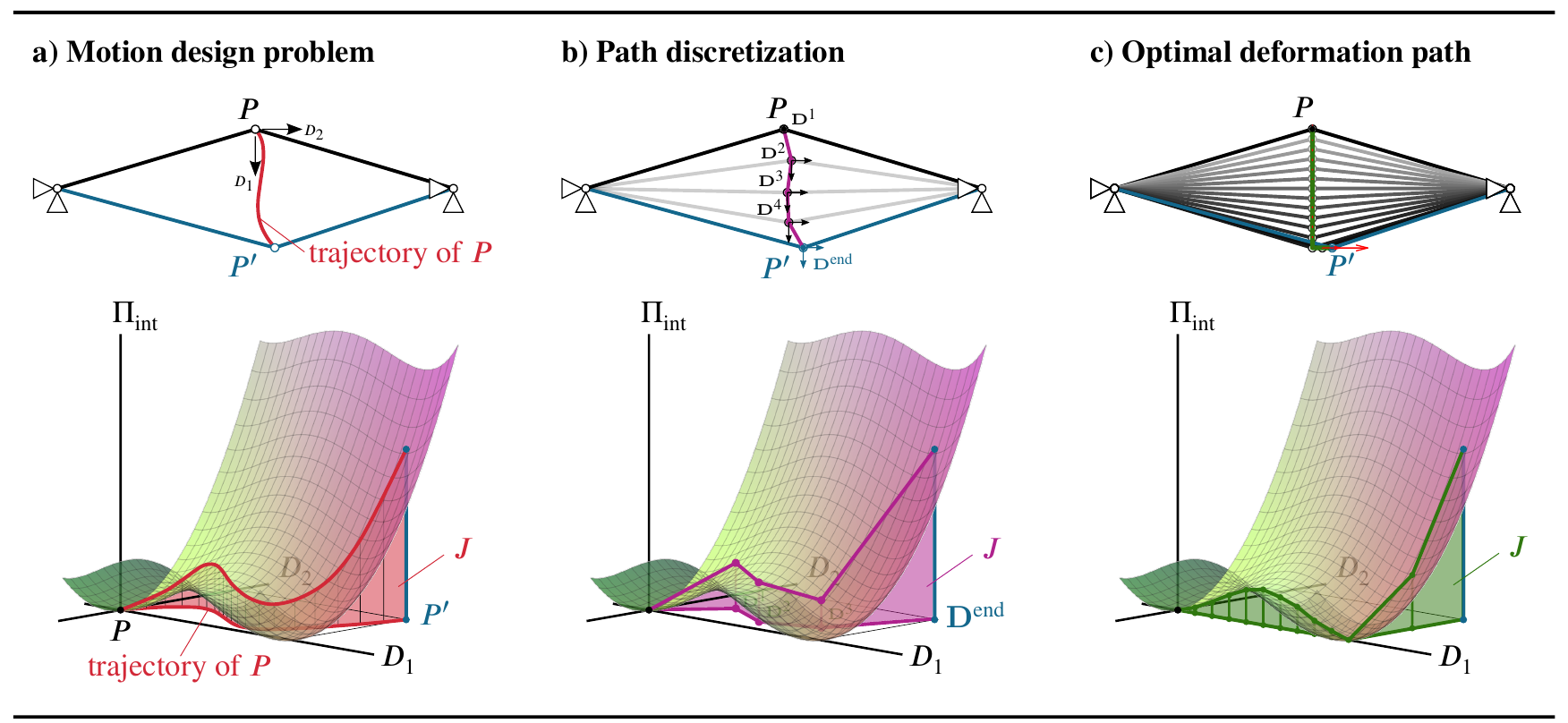}
\caption{Motion design on the example of a two bar truss and a snap-through problem.}
\label{fig:motion_design}
\end{figure}

A normalized arc length $\sb$ for a clear specification of the path integration bounds and a mapping parameter $s_\Ru = \fracdif{s}{\sb}$ are introduced. The first variation of $J$ is then
\begin{align}
\de J & = \int_0^1 \bigg[ \int_\OM \de \BE^\RT \BC \BE \dO s_\Ru + \int_\OM \frac 12 \BE^\RT \BC\BE \dO \de s_\Ru \bigg] \dsb  = 0 .
\label{eq:variation}
\end{align}
With the usual spatial discretization of the domain $\OM$ with finite elements, the variation can be rewritten as
\begin{align}
\de J & = \int_0^1 \bigg[ (\de \Bd^s)^\RT \BF_\Rint^s s_\Ru^s + (\de\Bd_{,s}^s)^\RT \Bs_\Ru^s \PI_\Rint^s \bigg] \dsb = 0,
\end{align}
where the internal energy $\PI_\Rint^s$ as well as the global vector of internal forces $\BF_\Rint^s$ can be identified.
Both quantities are still continuous with respect to the path parameter $s$, which is indicated by the superscript $(\;)^s$. 
In order to solve this equation, a second discretization, namely a discretization of the motion path from the initial configuration to a prescribed deformed target configuration with $\bar{n}_{\Rele}$ elements is introduced
\begin{align}
\int_0^1 (\ldots) \dsb = \sum_{{\bar{e}}=1}^{\bar{n}_{\Rele}} \int_{s^{\bar{e}}} (\ldots) \dsb^{\bar{e}} .
\end{align}
There is a formal analogy to space-time finite elements. However, the corresponding variable $s$ does not comply with time and the underlying functional $J$ is not obtained from the equations of motion.
Successive linearization and discretization of the motion path yields a linearized algebraic system of equations for motion design problems
\begin{align}
\BK_{\Rm\Rd} \DE \bar{\BD} & = -\BR_{\Rm\Rd} \,,
\label{eq:system_equation}
\end{align}
with the definitions
\begin{align}
\BK_{\Rm\Rd} & = \sum_{{\bar{e}}=1}^{\bar{n}_{\Rele}} \int_{\sb^{\bar{e}}} \Big( \bar{\BN}^\RT \BK_\RT s_\Ru \bar{\BN} + \bar{\BN}^\RT \BF_\Rint\Bs_\Ru \bar{\BN}_{,s} + \bar{\BN}_{,s}^\RT \Bs_\Ru \BF_\Rint \bar{\BN} + \bar{\BN}_{,s}^\RT \BS_\Ru \PI_\Rint \bar{\BN}_{,s} \Big) \dsb^{\bar{e}}, \\
\BR_{\Rm\Rd} & = \sum_{{\bar{e}}=1}^{\bar{n}_{\Rele}} \int_{\sb^{\bar{e}}} \Big( \bar{\BN}^\RT \BF_\Rint s_\Ru +
\bar{\BN}_{,s}^\RT \Bs_\Ru \PI_\Rint \Big) \dsb^{\bar{e}} .
\end{align}
$\bar{\BN}$ represents the matrix of shape functions for path discretization and can contain any type of function, i.e., Lagrange polynomials as well as spline functions. The vector of the total displacements
\begin{align}
\bar{\BD} & = \begin{bmatrix} \BD^{1} \\ \BD^{2} \\ \vdots \\ \BD^{\bar{k}} \\ \vdots \\ \BD^{\bar{n}_{\text{node}}} \end{bmatrix}
\end{align}
consists of $\bar{n}_\text{node}$ subvectors $\BD^s$ and gathers the displacements of every degree of freedom in every single configuration throughout the deformation process, where $\bar{n}_\text{node}$ denotes the number of nodes in the path discretization.
%This includes the initial reference configuration, deformed intermediate configurations and the (at least partly) prescribed deformed end configuration.
Every intermediate configuration corresponds to one node $\bar{k}$ of the path discretization. Thus, the shape functions in $\bar{\BN}$ serve as interpolation between the configurations. The total number of degrees of freedom is therefore
\begin{align}
\bar{n}_\Rdof = \bar{n}_{\text{node}} \cdot n_\Rdof.
\end{align}
The entire problem is solved monolithically by an iterative solution of the linearized system of equations~(\ref{eq:system_equation}). However, the problem might become ill-posed, as particular displacement trajectories may be representable by various different solution vectors. A simple example for this is a straight line, which can be represented by any sequence of straight steps with different size. To guarantee uniqueness of the solution, the evolution of at least one displacement degree of freedom has to be controlled throughout the entire deformation process. The entire derivation and further explanations are given in~\cite{sachse_motion_2019}. 

The following example briefly demonstrates the application of motion design. The initial geometry is a shallow, discrete two-bar truss, shown in black in Figure~\ref{fig:motion_design}a. It ought to be deformed to the target configuration shown in blue. The deformation path between the two configurations is supposed to minimize the integrated internal energy, defined as cost of deformation. During the deformation process, the point~$P$ follows a yet unknown trajectory (generically shown in red) until it arrives at its end position~$P'$. This trajectory and the corresponding internal energy can be identified on the diagram in Figure~\ref{fig:motion_design}a. The area of the spanned surface represents the value of the functional and is to be minimized by the resulting trajectory from motion design. The principle of path discretization is illustrated by four linear path elements in Figure~\ref{fig:motion_design}b. 

In our numerical experiment, however, actually 14 linear elements are used to achieve a better approximation.
Iterative solution of the system of equations in eq.~(\ref{eq:system_equation}) provides the entire deformation path in one go. The result is shown in Figure~\ref{fig:motion_design}b.
%The area of the resulting spanned surface by the optimized trajectory represents the minimized value of the functional.
It turns out that, for the given objective function $J$, it is optimal to first enforce a vertical snap through, followed by a horizontal movement towards the location of $P'$.

In order to technically realize the obtained deformation, forces are required. These external forces are obtained in a post-processing step after solving the motion design problem. From the displacement vectors $\BD^s$ of the optimized motion, the discrete internal forces can be computed for every configuration. The required external (actuation) forces are then obtained from equilibrium.
%Therefore, the goal of motion design is to calculate how non-zero forces evolve independently from each other during the deformation process to enable an optimized deformation path. 

%%%%%%%%%%%%%%%%%%%%%%%%%%%%%%%%%%%%%%%%%%%%%%%%%%%%%%%%%%%%%%%%%

\section{Constraints for prescribed sets of actuation forces}
\label{sec:con_load_cases}

\subsection{General concept}

\begin{figure}[h]
\centering
%\includesvg{fig_system_lc}
\hspace*{-4mm}
\includegraphics[width=15.0cm]{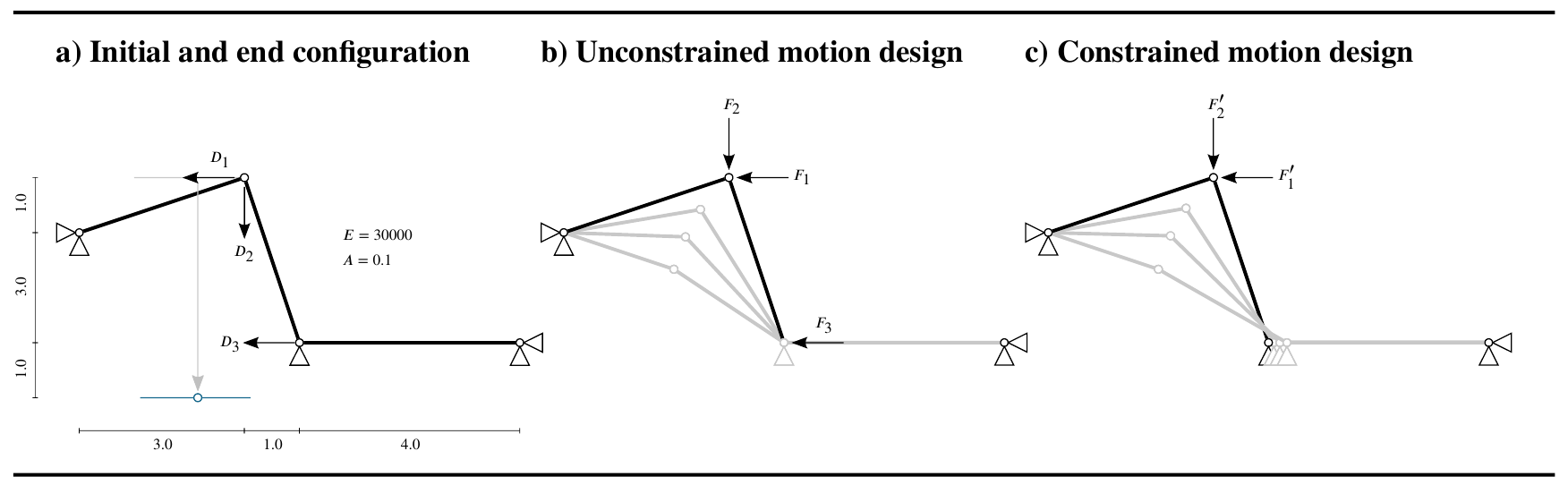}
\caption{Illustrating example for unconstrained and constrained motion design.}
\label{fig:system_lc}
\end{figure}

The motion design method from~\cite{sachse_motion_2019} relies on the hypothesis that discrete actuation forces can potentially be applied to every degree of freedom. This is illustrated in the simple truss system in Figure~\ref{fig:system_lc}a. The initial configuration is shown in black. In this example, the target configuration is not fully prescribed, but only the vertical displacement $D_2$ of the second node is supposed to take a specific value. In an unconstrained motion design, a motion path is found on the basis that all forces $F_1$, $F_2$ and $F_3$ are potentially non-zero and thus available for actuation, as indicated in Figure~\ref{fig:system_lc}b.
%This represents a huge restriction for real structures as usually a limited amount of loads is available to realize the motion. 

In constrained motion design, these forces are subject to conditions, which can be quite general. Practically, these conditions emanate from considerations concerning technical feasibility. In the simplest case, such a condition can be that a certain force is zero, i.e., there is no external actuator available to apply a force in the corresponding direction. 
%In constrained motion design is to calculate how the prescribed non-zero forces need to evolve independently to follow a resulting optimized deformation path. Furthermore, these non-zero forces may also incorporate dependencies as it is the case in whole load cases such as line loads or surface loads. Here, a scaling factor of the load case controls all associated discrete forces in the same manner. Applying such constraints also affects the end geometry, which is also discussed in this chapter.
In Figure~\ref{fig:system_lc}c, this is represented by a system, in which only two forces $F'_1$ and $F'_2$ can be applied to the structure.
%These applied forces on the structure are herein denoted with a dash.
$F_3$ has to be zero during the entire motion. This also affects the end geometry, which is further discussed in the following. 

\subsection{Equality constraints for uncontrolled degrees of freedom}
%\label{sec:con_load_cases}

The fact that the desired motion has to be realized with a limited set of actuation forces represents a constraint within the motion design problem. The main functional and objective remain the same.
%Following the method presented in~\cite{sachse_motion_2019}, the integrated internal energy over the motion is also used here as a functional. However, it can also be replaced.
Generally, the task of motion design with equality constraints can be formulated as: Minimize the functional $J$ subject to constraints $\Bg = \B0$.

The constraints are enforced by an extension of the functional. Here, the Lagrange multiplier method is applied. However, other methods can be used, e.g., the penalty method or the Augmented Lagrange method. Within the Lagrange multiplier method, additional variables, the Lagrange multipliers $\Bla$, are introduced for each constraint. It then follows for the extended functional
\begin{align}
\bar{J} = J + \Bla^\RT \Bg .
\label{eq:func_ext_lc}
\end{align}
According to the constraint, external actuation forces can only act in the direction of specific degrees of freedom. For all other degrees of freedom, the corresponding internal forces have to be zero due to equilibrium.
%Following the motion design approach, these can freely evolve and adjust during the deformation process. On the other hand, the internal forces at the unloaded degrees of freedom are supposed to continually remain zero.
This condition can be expressed as a set of equality constraints on the motion. 

As the described constraints must hold for the entire motion, they are introduced for each configuration during the deformation process. They therefore represent pointwise constraints, which are not included in the integral of the main functional and can be formally expressed as
\begin{align}
\Bg & = \begin{bmatrix} \BF^{1} \\ \BF^{2} \\ \vdots \\ \BF^{\bar{k}} \\ \vdots \\ \BF^{\bar{n}_{\text{node}}} \end{bmatrix} = \B0 .
\end{align}
Here, the vectors $\BF^{\bar{k}}$ gather the internal forces of only those degrees of freedom that are not subject to actuation forces. \footnote{When using a path discretization with an approximation by B-spline shape functions, these constraints should not be imposed on the control points but on the deformation path itself. For this reason, the constraints are enforced at collocation points, which are located on the path. In this case, Greville points are used.}

For the system presented in Figure~\ref{fig:system_lc} and a motion path discretization with  linear path elements, the internal force at degree of freedom number~3 must be zero in every and five path increment. Consequently, the vector of constraints can be written as
\begin{align}
\Bg & = \begin{bmatrix} F^{2}_3 & F^{3}_3 & F^{4}_3 & F^{5}_3 \end{bmatrix}^\RT ,
\end{align}
i.e., the internal force at degree of freedom number~3 must be zero in every path increment. The internal force $F_3^1$ of the initial configuration is omitted as the initial configuration is always assumed to be load-free.

If distributed line loads or surface loads are applied as actuation forces, the discrete forces of the individual degrees of freedom share the same scaling factor. In such cases, a master degree of freedom is selected and the associated forces are coupled to this load value via an equality constraint.

After linearization, the algebraic system of equations of the constrained problem reads 
\begin{align}
\begin{bmatrix}
\BK_{\Rm\Rd} + \la_i \BH_i & \BG^\RT \\ \BG & \B0
\end{bmatrix}
\begin{bmatrix}
 \DE \bar{\BD} \\ \DE \Bla
\end{bmatrix} = - \begin{bmatrix}
\BR_{\Rm\Rd} + \Bla^\RT \BG \\ \Bg
 \end{bmatrix} .
 \label{eq:system_eq_ext}
\end{align}
Here, the matrix $\BG$ contains the first derivatives of all constraints with respect to the degrees of freedom $\bar{\BD}$,
\begin{align}
\BG = \fracpt{\Bg}{\bar{\BD}},
\end{align}
the Hessian $\BH_i$ contains the second derivatives of the $i^\text{th}$ constraint,
\begin{align}
\BH_i = \fracpt{g_i}{\bar{\BD}\partial\bar{\BD}}.
\end{align}
 This means that the constraints and thus the corresponding internal forces $\BF^{\bar{k}}$, need to be differentiated twice with respect to all degrees of freedom. 
The internal force at a degree of freedom $j$ in the load configuration $\bar{k}$ only depends on the degrees of freedom at the associated path node. Moreover, the tangent stiffness matrix $\BK_\RT^{\bar{k}}$ at this path node $\bar{k}$ already incorporates the first derivative of the internal forces with respect to the relevant degrees of freedom. Accordingly, the derivative of one component $F^{\bar{k}}_j$ of the total internal force vector $\BF^{\bar{k}}$ is provided by the associated column $\BK_j^{\bar{k}}$ of the stiffness matrix. Consequently, the vector $\BG_i$ of first derivatives of the $i^\text{th}$~constraint $g_i$ follows as
\begin{align}
\BG_i = \fracpt{g_i}{\BD^{\bar{k}}} = \fracpt{F_j^{\bar{k}}}{\BD^{\bar{k}}} = \BK_j^{\bar{k}} ,
\end{align}
where $j$ is the degree of freedom that corresponds to the constraint $g_i$. 
Additionally, the second derivative of the $i^{\text{th}}$ constraint is required for the Hessian matrix $\BH_i$. In this case, this represents the derivative of the associated column of the stiffness matrix. Usually, the relevant column of the stiffness matrix $\BK_j$ has to be differentiated with respect to each degree of freedom. This yields for the Hessian matrix of the $i^\text{th}$~constraint
\begin{align}
\BH_i = \fracpt{\BK_j^{\bar{k}}}{\BD^{\bar{k}}} \, .
\label{eq:H_diff_1}
\end{align}
However, the analytical derivative of the tangent stiffness matrix can only be computed in special cases, such that numerical differentiation has to be applied. For classical forward, backward or central difference schemes the choice of the perturbation in order to minimize both the truncation error and the elimination error is not trivial. Therefore, here complex step differentiation, as presented in~\cite{martins_complex-step_2003} is used. It allows for arbitrarily small perturbations because there is no elimination error and thus accuracy up to machine precision is always guaranteed.

Nevertheless, numerical differentiation with respect to each individual degree of freedom is numerically expensive, as it involves repeated computation of the stiffness matrix for each direction of differentiation.
However, because derivatives commute and the tangent stiffness matrix is already calculated through displacement derivatives, the derivative of the $j^{\text{th}}$ column with respect to the $m^{\text{th}}$ degree of freedom equals the derivative of the $m^{\text{th}}$ column with respect to the $j^{\text{th}}$ degree of freedom, i.e., $\fracpt{\BK_j}{D_m} = \fracpt{\BK_m}{D_j}$. Therefore, the required differentiation with respect to column $j$ of the stiffness matrix can be extracted from the derivative of the total stiffness matrix by a scalar product with a unit vector $\BI_j$ in direction of $j$
\begin{align}
\BH_i = \fracpt{\BK_j^{\bar{k}}}{\BD^{\bar{k}}} = \fracpt{\BK^{\bar{k}}}{\BD^{\bar{k}}} \cdot \BI_j .
\label{eq:H_diff}
\end{align}
Thus, the derivative in eq.~(\ref{eq:H_diff_1}) is transformed to a directional derivative in eq.~(\ref{eq:H_diff}) and the advantage of using numerical differentiation regarding numerical efficiency becomes apparent. Instead of differentiating a single column of the stiffness matrix with respect to all degrees of freedom and consequently evaluating it as many times for numerical differentiation, the whole stiffness matrix now only has to be calculated one additional time in the direction of $\BI_j$. This approach also has major advantages regarding implementation into existing finite element software, as usually the stiffness matrix is calculated in an entirety and not only distinct columns.
%In order to increase accuracy, the numerical derivative is determined with the help of the complex step approximation according to \cite{martins_complex-step_2003}. As already explained, the deviation is executed in the complex direction and therefore, the first derivative can be calculated numerically and, more particularly, exactly.

The derivatives of the system in Figure~\ref{fig:system_lc} for the first ($i=1$) constraint $F^{2}_3 = 0$ are exemplarily presented in the following. In this case, there is only one relevant degree of freedom, namely $j=3$. The first derivative can be extracted from the total stiffness matrix of the current load configuration $\bar{k}=2$, where it represents the third column
\begin{align}
\BG_1 = \fracpt{F^{2}_3}{\BD^2} = \BK^2_3 .
\end{align}
The Hessian matrix contains the whole stiffness matrix of the second configuration and can be calculated with the help of a directional derivative
\begin{align}
\BH_1 = \fracpt{\BK^2}{\BD^2} \cdot \BI_3.
\end{align}
With these derivatives, the extended system of equations from eq.~(\ref{eq:system_eq_ext}) can be generated and solved. Thus, an optimized motion can be found that is realized exclusively by the defined actuation forces. The result is the deformation path and the development of the actuation force amplitudes during the motion.

\subsection{Restrictions for the prescribed target geometry}
\label{sec:con_end_geometry}

In the basic motion design method, either the entire target geometry or only parts of it have to be prescribed. This is similar for constrained motion design, but with one important restriction: %The end geometry should be able to be reached only with the allowed load cases. 
As equilibrium has to be fulfilled using only the available actuation forces throughout the entire motion, this must also be the case for the prescribed target geometry. Otherwise, the problem is not well-posed.

Two different cases have to be distinguished, depending on the number of prescribed target displacement values:

\begin{itemize}
	\item If the entire target geometry is prescribed, an optimization is carried out prior to the motion design process. The objective function to be minimized is the difference between the target geometry and the current geometry with the constraint that the internal forces in direction of non-actuated degrees of freedom must be equal to zero.
	\item If only a subset of degrees of freedom is prescribed, the rest of the geometry can adjust freely to meet the condition of equilibrium with the available actuation forces. The maximum number of prescribed values of displacement degrees of freedom depends on the number of independent actuation forces. It is not possible to prescribe more displacement values than this value, whereas fewer are generally possible. In cases, where it is not obvious, whether an equilibrium configuration is possible, the same optimization procedure as described above can be applied.
\end{itemize}

\subsection{Solution and interpretation of the results}

\begin{figure}[b]
\centering\small
%\includesvg{fig_system_lc_sol}
\hspace*{-4mm}
\includegraphics[width=15.0cm]{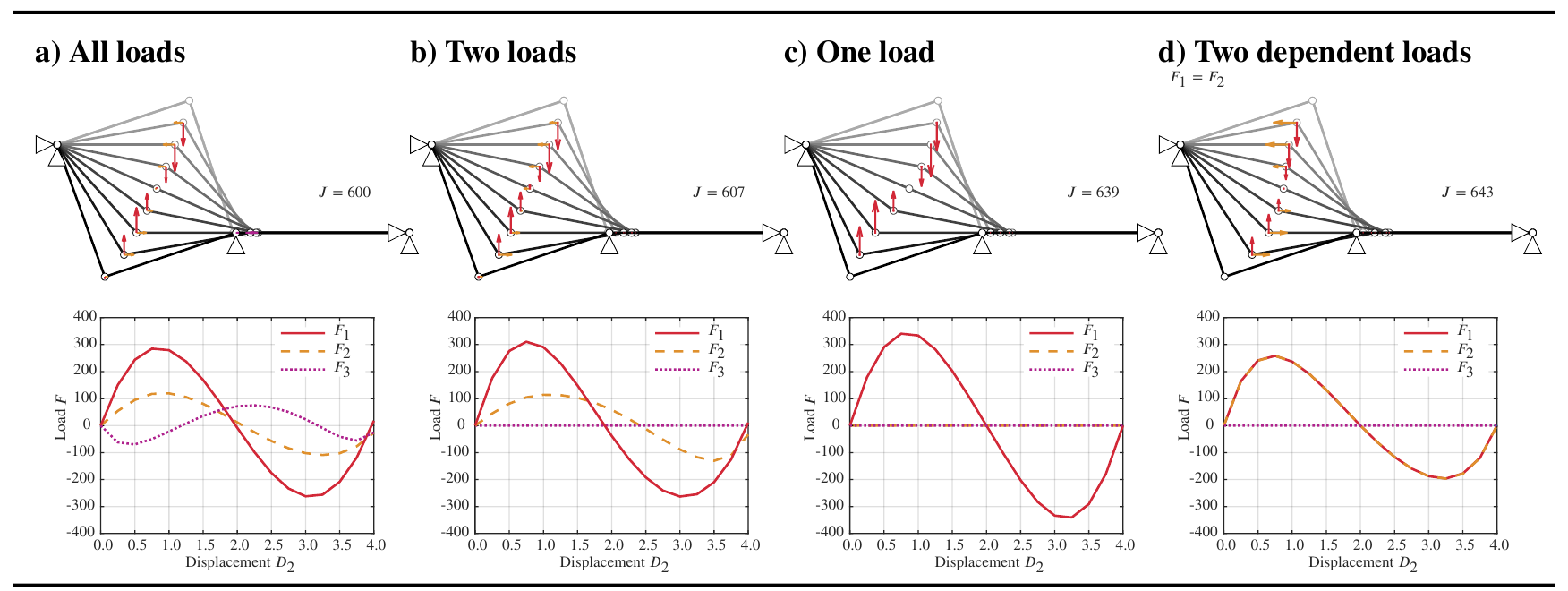}
\caption{Solutions of the illustrating example with different numbers and types of load cases.}
\label{fig:system_lc_sol}
\end{figure}

Taking into account the described restrictions for the target geometry, the extended system of equations~(\ref{eq:system_eq_ext}) can be solved. In doing so, an optimized motion path is obtained that considers the given constraints of enforcing an equilibrium state with only the available actuation forces. The results for the example in Figure~\ref{fig:system_lc} with a varying number of actuation forces are summarized in Figure~\ref{fig:system_lc_sol}.

If forces can be applied on every degree of freedom, no constraints are enforced and the solution is identical to the result from an unconstrained motion design. This is illustrated in Figure~\ref{fig:system_lc_sol}a. It can be seen in the force-displacement curves that all the three actuation forces take non-zero values during the deformation process to realize the calculated optimal deformation path. This motion yields a minimized functional value of $J=600$.

If one load is suspended (Figure~\ref{fig:system_lc_sol}b), this optimal deformation path cannot be followed anymore since all three possible point loads are required for keeping an equilibrium state in the deformed configurations. Therefore, a different motion is found that is enabled with the remaining forces. Moreover, as only the vertical displacement of the target geometry is prescribed, the horizontal displacement adapts such that the additional constraint is met, eventually resulting in a different final configuration. The minimum of the functional increases to $J=607$. If only one force, or one dependent couple of forces $F_1 = F_2$ (see Figure~\ref{fig:system_lc_sol}c/d) is allowed for actuation, the result is identical to an equilibrium path obtained by a non-linear analysis with either the arc length method or a displacement controlled algorithm. The motion does not represent a ``design'' in this case because there is no other possible equilibrium path (neglecting here the theoretical possibility of bifurcation points and corresponding secondary paths).

In order to verify the applicability of the presented method, numerical experiments are presented in the next section. Among others, problems from the~\cite{sachse_motion_2019} are reused to demonstrate the correlation between unconstrained and constrained motion design with restrictions concerning the actuation forces.

\subsection{Numerical experiments for motion design with constrained actuation}

\subsubsection{Motion design with multiple snap-through processes}
\begin{figure}[b]\centering\small
%\includesvg{fig_example_lc_truss}
\hspace*{-4mm}
\includegraphics[width=15.0cm]{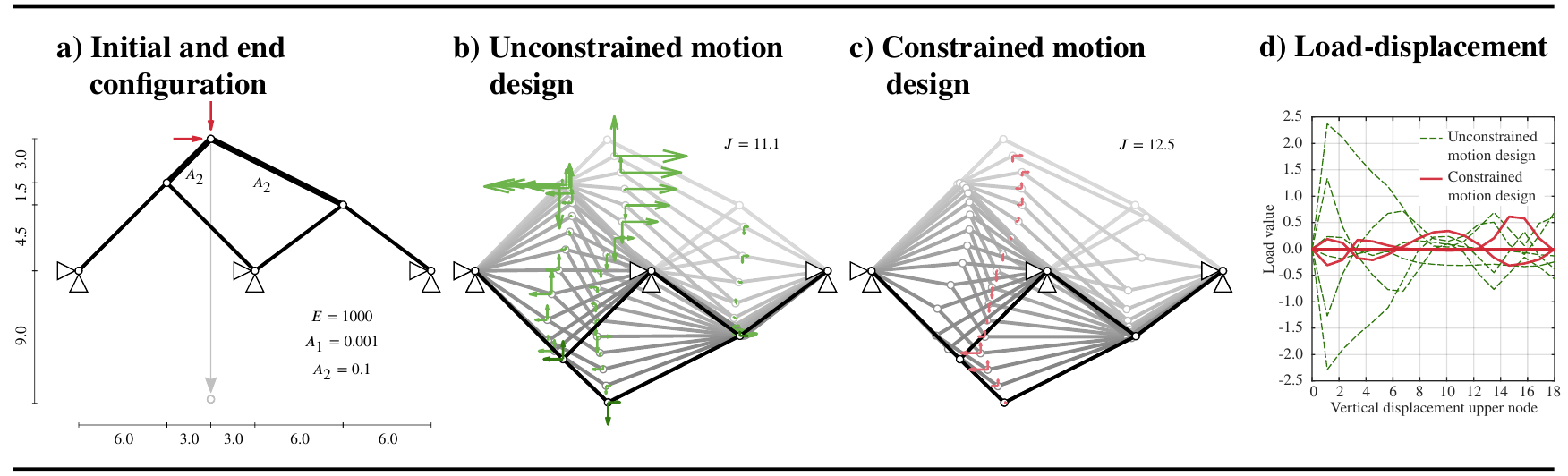}
\caption{Combination of multiple snap-throughs with unconstrained and constrained motion design.}
\label{fig:example_lc_truss}
\end{figure}

\phantom{test}\\[2mm]
First, an example is presented that was already previously studied in~\cite{sachse_motion_2019}, in order to demonstrate the difference between unconstrained and constrained motion design. Three two-bar trusses are combined into one truss structure. The members of the upper two-bar truss have a larger cross-sectional area and therefore higher stiffness than the lower ones, by which it is supported (see Figure~\ref{fig:example_lc_truss}a. The vertical displacement of the upper node is prescribed in the target configuration. The result of an unconstrained motion design procedure with a path discretization with 16 linear elements is shown in Figure~\ref{fig:example_lc_truss}b. As already described in~\cite{sachse_motion_2019}, a naive linear interpolation of all displacement values between the initial and target values as ``predictor motion'' yields a high value of $J=440$ for the quality functional. With unconstrained motion design this reduces to $J=11.1$. Figure~\ref{fig:example_lc_truss}b visualizes the corresponding motion along with the required actuation forces. To realize this optimal motion, actuation forces need to be applied on all six degrees of freedom, which can be seen in the load-displacement curves (green) in~Figure~\ref{fig:example_lc_truss}d.

For a constrained motion design, only two actuation forces, namely a vertical and a horizontal one at the upper node are allowed. This results in four constraints per load configuration, which enforce the remaining internal forces at the uncontrolled degrees of freedom to be zero. Solving the extended system of equations leads to the modified motion illustrated in Figure~\ref{fig:example_lc_truss}c.
The end geometries of the unconstrained and constrained motion design variants differ slightly from each other and from the mirrored geometry, but the vertical displacement of the upper node coincides in both cases, as this represents the prescribed target value. Furthermore, it can be seen that retaining the last equilibrium state requires lower forces in constrained motion design. 
Since the members of the upper two-bar truss has a much larger cross-sectional area than those of the two lower two-bar trusses, the small deviation from the mirrored geometry has a significant influence on the magnitude of the forces, recovered from the displacements for the case of unconstrained motion design. In the constrained motion, however, forces are allowed to only act on one single node. % nodes connected to elements with a large cross-sectional area. 
This reduces the described effect on the loads attributable to the not perfectly matched mirrored end geometry.
Due to the additional constraints, the value of the functional increases to $J=12.5$.
The two resulting motions from unconstrained and constrained motion design resemble each other, especially regarding the overall motion pattern with the consecutive lateral snap-through processes.
However, the magnitude of the required forces differ significantly; the forces required to follow the constrained motion are much lower (see Figure~\ref{fig:example_lc_truss}d).
%To sum up, the restriction to realize the motion with only two forces instead of all six possible point loads leads to a slight increase of the functional value, while achieving a similar appearance of the motion in this particular scenario. Furthermore, the required absolute force values become much smaller even though the number of loads is decreased by four.

\subsubsection{Snap-through of a shallow arc}
\begin{figure}[t]
\centering\small
\hspace*{-4mm}
\includegraphics[width=15.0cm]{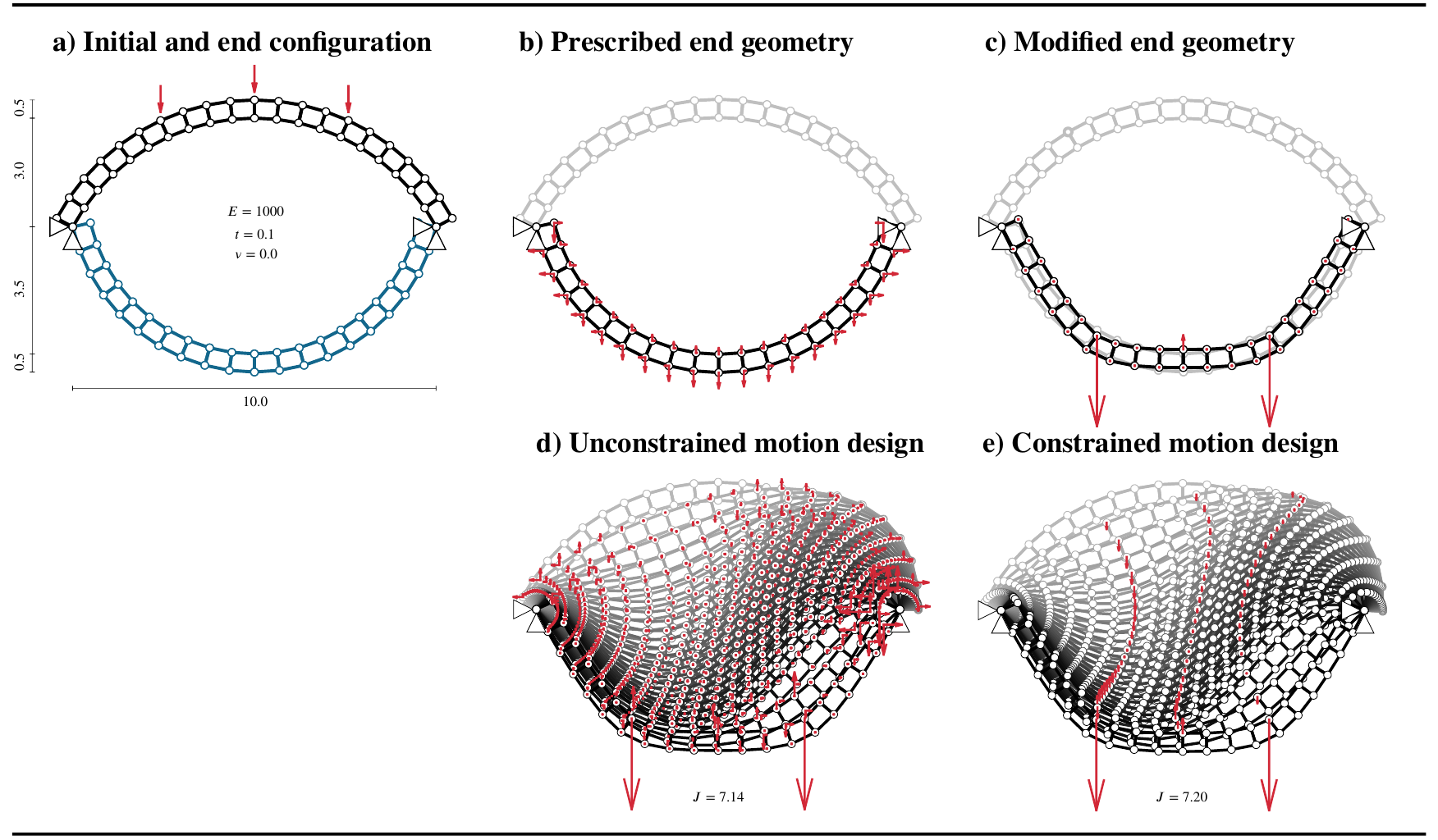}
%\includesvg{fig_example_lc_arc}
\caption{Constrained motion design of a shallow arc with three point loads.} 
\label{fig:example_lc_arc}
\end{figure}

\phantom{test}\\[2mm]
The next problem is a two-dimensional plane stress problem in the shape of a shallow circular arc. It is discretized with displacement-based quadrilateral finite elements. The target geometry, shown in blue in~Figure~\ref{fig:example_lc_arc}a is defined to be a circular arc that resembles -- but is not -- an equilibrium configuration after snap-through. Only three vertical actuation forces are applied, shown in red.
%The end configuration differs from the previously defined end configuration in \cite{sachse_motion_2019} to better illustrate the effect of the applied loads. 

Since the target geometry is not in equilibrium, point forces have to be applied at every degree of freedom for its realization, see Figure~\ref{fig:example_lc_arc}b. This contradicts the constraint that the motion ought to be realized with only three actuation forces. Therefore, the target geometry has to be modified as explained in Section~\ref{sec:con_end_geometry}. An optimization procedure is carried out, in which the target displacements are the design variables and the displacement difference between the prescribed and the current geometry is minimized. The optimization problem is subject to the constraint that equilibrium is obtained with the three available point forces only. Figure~\ref{fig:example_lc_arc}c illustrates that the resulting geometry differs noticeably from the initially prescribed geometry.

This newly obtained geometry is the target configuration for the motion design problems solved next. The deformation path is discretized by 24 linear elements. An unconstrained motion design yields the motion in Figure~\ref{fig:example_lc_arc}d with a functional value of $J=7.14$. It turns out that a non-symmetric snap-through of the arc is beneficial for a low value of $J$. Although the target geometry can be equilibrated with the three prescribed forces only, non-zero forces are applied at the other degrees of freedom throughout the motion to realize the optimal deformation path in unconstrained motion design.

This result is obviously different than the one from constrained motion design, which is shown in Figure~\ref{fig:example_lc_arc}e. Here, only the three admissible actuation forces are applied during the entire deformation process. Due to this constraint, the functional value increases slightly to $J=7.20$.
%
%\textcolor{red}{\sout{This result from motion design is compared to a classical non-linear analysis in Figure~\ref{fig:example_lc_arc}f, where the loads from the end configuration are applied. The solution is calculated with the help of the arc length control for path-following within 24 steps. As all three forces are increased and decreased uniformly, a symmetric snap-through appears. The resulting deformation path varies significantly from the one obtained by motion design. The functional takes a value of $J=8.40$, which is higher than those from both unconstrained and constrained motion design.}} 
This resulting optimal motion could not be followed within a non-linear analyses, in which the relation between the three different actuation forces is locked.
Thus, even though the constraints increase the cost of deformation, it is still lower than for the conventional approach of a non-linear analysis. This is the consequence of enabling an independent evolution of the different actuation forces throughout the deformation process.

%%%%%%%%%%%%%%%%%%%%%%%%%%%%%%%%%%%%%%%%%%%%%%%%%%%%%%%%%%%%%%%%%

\section{Discrete actuator elements for internal actuation}
\label{sec:con_actors}

\subsection{Actuator element formulation}

\begin{figure}[h]\centering%\small
%\includesvg{fig_actuator}
\hspace*{-4mm}
\includegraphics[width=15.0cm]{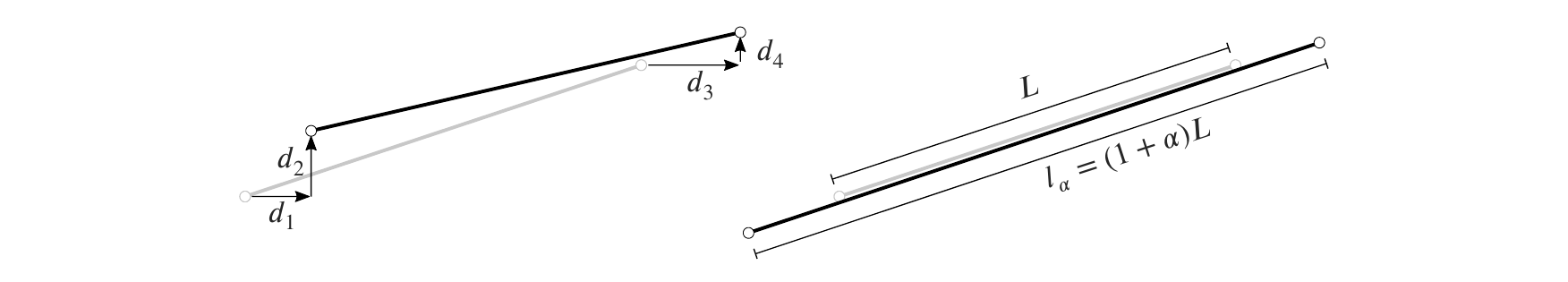}
\caption{Parameter in the actuator element formulation.}
\label{fig:actuator}
\end{figure}

As an alternative to applying external forces, actuation of structures can also be accomplished via actuator elements. Simple, linear actuators in truss structures can contract or expand and thus enforce a deformation of the entire structure or parts of it. To include such elements into the motion design method, a new actuator element formulation is introduced. The basis of the actuator element is a standard truss element with a vertical and horizontal displacement degree of freedom at each node (cf. Figure~\ref{fig:actuator} (left)). The actuator element, however, also allows for a pure and independent elongation or contraction. This is mapped by an additional parameter $\al$, which represents a factor for the targeted actuator element elongation, as can be seen in Figure~\ref{fig:actuator} (right).

In an actuator element, the total strain $E_{\Rd}$ consists of two parts: the elastic strain $E_{\Rel}$ and the strain from the actuation $E_{\al}$
\begin{align}
E_{\Rd} = E_{\Rel} + E_\al.
\end{align}
The actuation strain $E_\al$ is determined with the additional elongation parameter $\al$ and the total strain $E_\Rd$ is obtained from the nodal degrees of freedom via the kinematic equations. This allows evaluation of the elastic strain and eventually the normal stress from the material law. Applying the linear elastic St.~Venant-Kirchhoff material law, the internal energy is obtained as
\begin{align}
\PI_{\Rint,\Rel} = \int_\OM E_{\Rel}^\RT \BC E_{\Rel} \dO = \int_\OM (E_{\Rd} - E_\al)^\RT \BC (E_{\Rd} - E_\al) \dO .
\end{align}
From this, the internal forces can be derived, yielding for a single actuator element 
\begin{align}
\Bf_{\Rint} = \int_{\OM_e} \BB^\RT \BC (E_{\Rd} - E_\al) \dO .
\label{eq:f_int_act}
\end{align}
The strain-displacement matrix $\BB$ is defined as the derivative of the elastic strain with respect to the unknown parameters and is divided into two parts: the derivative with respect to the local displacement degrees of freedom $\Bd$ and the derivative with respect to the elongation parameter $\al$. This results in
\begin{align}
\BB = \begin{bmatrix} \BB_\Rd \\ B_\al \end{bmatrix} 
\qquad \text{with} \qquad 
\BB_\Rd = \fracpt{E_\Rd}{\Bd}
\qquad \text{and} \qquad
B_\al & = -\fracpt{E_\al}{\al}.
\end{align}
Consistent linearization of the global residual equation for equilibrium $\BR = \BF_\Rint - \BF_\Rext$ with the global vector of internal forces $\BF_\Rint$ and external forces $\BF_\Rext$ yields the stiffness matrix. The local stiffness matrix for the actuator element follows as
\begin{align}
\Bk_{\Ract} = 
\begin{bmatrix}
\BB_{\Rd,\Bd}^\RT \BC E_\Rel + \BB_\Rd^\RT \BC \BB_\Rd  & B_\al \BC \BB_\Rd \\
\BB_\Rd^\RT \BC B_\al & B_{\al,\al}^\RT \BC E_\Rel + B_\al^\RT \BC B_\al 
\end{bmatrix} 
= 
\begin{bmatrix}
\Bk_{\Rd\Rd}  & \Bk_{\Rd\al} \\
\Bk_{\Rd\al}^\RT & \Bk_{\al\al}
\end{bmatrix} .
\end{align}
It includes the usual tangent stiffness matrix $\Bk_{\Rd\Rd}$ of a truss element in the upper left corner. The global stiffness matrix can be obtained by standard assembly operations for $n_\Rele$ elements. These operations can be applied separately for the different stiffness components
\begin{align}
\BK_{\Rd\Rd} = \bigcup_{e=1}^{n_\Rele} \Bk_{\Rd\Rd} 
\qquad \qquad
\BK_{\Rd\al} = \bigcup_{e=1}^{n_\Rele} \Bk_{\Rd\al} 
\qquad \qquad
\BK_{\al\al} = \bigcup_{e=1}^{n_\Rele} \Bk_{\al\al} ,
\end{align}
which results in the global linearized system of equations for equilibrium
\begin{align}
\begin{bmatrix}
\BK_{\Rd\Rd}  & \BK_{\Rd\al} \\
\BK_{\Rd\al}^\RT & \BK_{\al\al}
\end{bmatrix}
\begin{bmatrix}
\DE \Bd \\ \DE \Bal
\end{bmatrix}
= + \BF_\Rint - \BF_\Rext .
\label{eq:system_act}
\end{align}

A prescribed elongation or contraction of an actuator can be treated as inhomogeneous Dirichlet boundary condition. The procedure can also be interpreted as an actuation with external forces, extracted from eq.~(\ref{eq:f_int_act}) as
\begin{align}
\Bf_{\Rint,\Ract} = - \int_{\OM_e} \BB^\RT \BC E_\al \dO .
\end{align}
%Thereupon, it is solved for the resulting displacements.

\begin{figure}[t]\centering%\small
%\includesvg{fig_system_act}
\hspace*{-4mm}
\includegraphics[width=15.0cm]{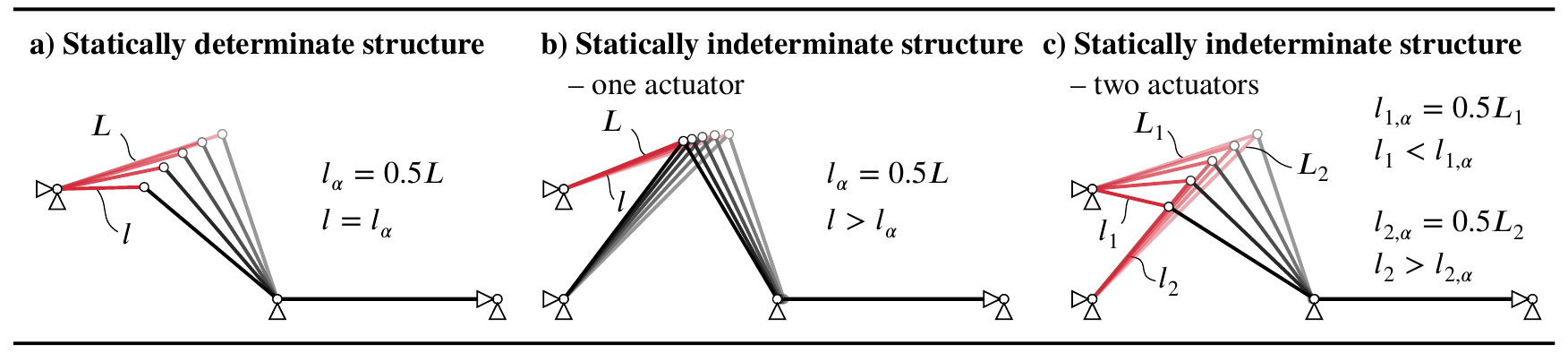}
\caption{Actuator length change as load case and the resulting deformation of a statically determinate and statically indeterminate structure with one and two actuator elements.}
\label{fig:system_act}
\end{figure}

The procedure is demonstrated in the following example, a statically determinate structure, which was already introduced in the previous section. Figure~\ref{fig:system_act}a shows the actuator element in red. A shortening by 50\% is prescribed.
%Thus, the elongation parameter for this element is set to the value of $\al = -0.5$. The other two elements represent regular truss elements.
With the system of equations in eq.~(\ref{eq:system_act}) and a load-controlled geometrically non-linear analysis, the resulting nodal displacements can be computed. The solution is shown in Figure~\ref{fig:system_act}a.
%It can be observed that only the actuator element experiences a length change while the other two bars undergo a kinematic deformation keeping their lengths.
%As one truss element is replaced by an actuator element in this statically determinate structure, there are no constraints when shortened.
Because the structure is statically determinate, a stress-free, purely kinematic motion is obtained.
%Due to that, the actuator element reaches exactly the targeted length change of 50\% and its actual length corresponds to half of the original length. However, such a constraint-free shortening of the actuator element is only possible to a certain extent and strongly depends on the structure.

Figure~\ref{fig:system_act}b shows the same system with an additional bar, such that it is now statically indeterminate by degree $n_\Rs = 1$. The same actuator element and the same target length change as before are used. The statical indeterminacy of the structure in combination with the position of the single actuator obstructs a contraint-free motion. Stresses are implied to the other truss elements and the actuator does not reach the targeted length change of 50\% when specifying $\al = -0.5$. The resulting deformed structure is shown in Figure~\ref{fig:system_act}b.

When a second actuator element replaces a regular truss element, a constraint-free length change as well as a kinematic motion are potentially possible. However, it is not possible with the given actuator combination and the prescribed elongation parameters of both $\al = -0.5$ as displayed in Figure~\ref{fig:system_act}c. The truss elements still perform nearly the kinematic motion. However, the length changes of both actuators do not exactly match this specific motion, creating constraints in the actuators and leading to different length changes than prescribed. This brief study shows that the number and location of actuators and the degree of static indeterminacy
%and their corresponding elongation parameters
are crucial for the feasibility of a constraint-free kinematic motion.

%These examples depict the expected structural behavior and deformation results and accordingly, the actuator element formulation presented here is suitable for structural analyses with such actuator elements.

\subsection{Motion design with actuator elements}

The introduced actuator element formulation allows an easy implementation into the motion design method. For this purpose, the elongation parameters $\al_i$ for every actuator element $i$ are discretized along the motion path, as it has been done with the displacement degrees of freedom. Thus, the elongation parameters $\al_i$, as part of the solution vector, are directly solved for while optimizing the motion. Again, this is the same process as for the displacement degrees of freedom. Therefore, the elongation parameters vary according to the optimized motion and their evolution throughout the deformation process represents an output of the motion design method.

This is again illustrated with the truss example from Figure~\ref{fig:system_lc}, prescribing the same vertical displacement $D_2$ of the second node, using a path discretization with twenty linear path elements. With the presented actuator element formulation, the basic motion design method already includes the actuation load case, i.e., without applying any external forces. In order to ensure that only the actuator is used to realize the optimized motion without any discrete point loads, external forces are constrained to be zero for every degree of freedom (see Chapter~\ref{sec:con_load_cases}).

\begin{figure}[t]\centering\small
%\includesvg{fig_system_act_md}
\hspace*{-4mm}
\includegraphics[width=15.0cm]{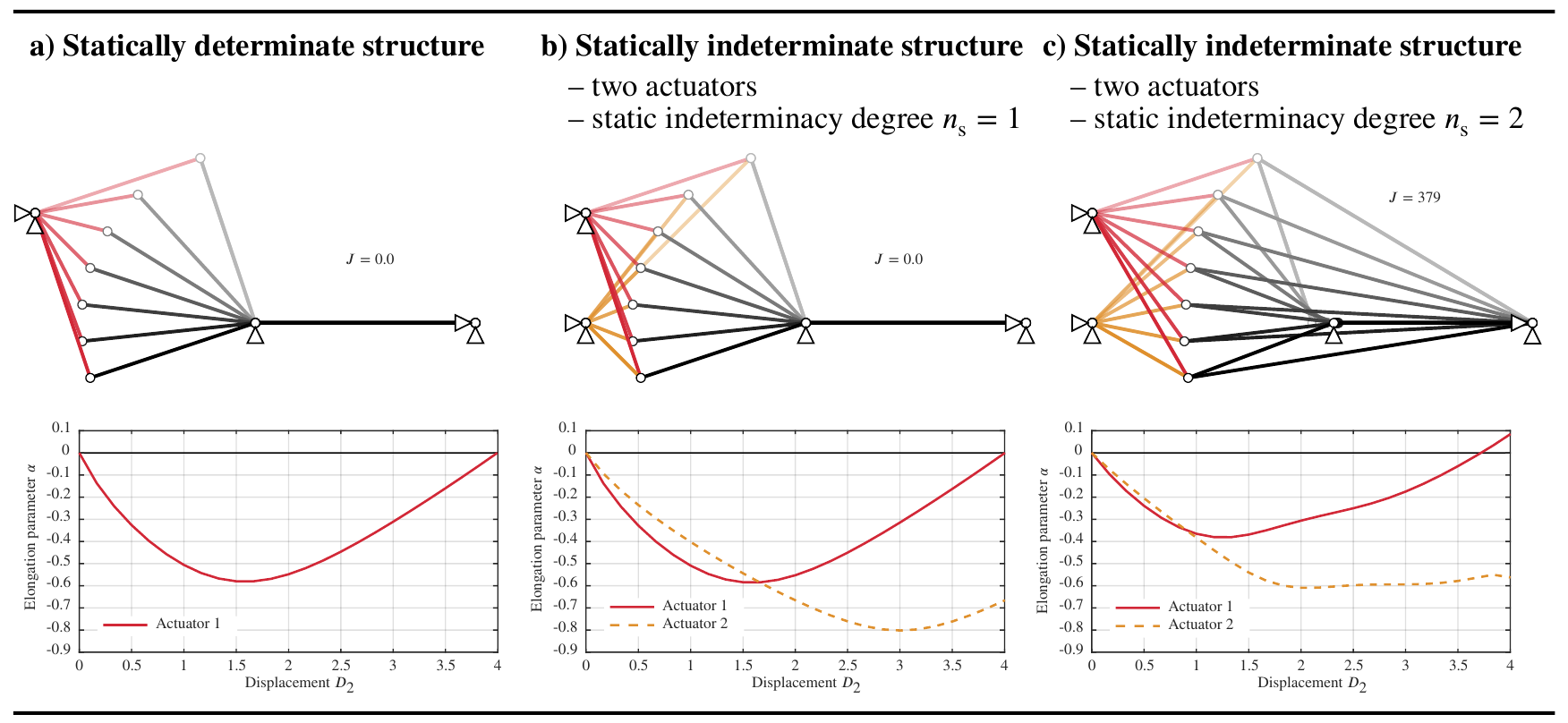}
\caption{Motion design with actuator elements.}
\label{fig:system_act_md}
\end{figure}

First, the statically determinate structure with one actuator element is investigated, cf.~Figure~\ref{fig:system_act_md}a. Here, a purely kinematic motion of the rest of the structure is obtained, which can be identified by the functional value being $J=0.0$.
%This motion minimizes the functional, which is defined as the integral of the elastic energy of the regular truss elements.
The most important output of motion design is the evolution of the elongation parameter throughout the deformation process. The actuator adapts to the optimized motion, as can be seen in the plot of the elongation parameter versus the displacement $D_2$ in Figure~\ref{fig:system_act_md}a (bottom). This can also be observed in the statically indeterminate structure (degree $n_\Rs = 1$) with two actuator elements in Figure~\ref{fig:system_act_md}b. In this case, a suitable combination of the elongation parameters of both actuator elements is found by the motion design method, such that a purely kinematic motion is obtained. Again, this results in a functional value of $J=0$. The motion is different from the one in Figure~\ref{fig:system_act}, because now the elongation parameters of both actuators may evolve independently.

Finally, a structure with a second additional truss element is studied (cf.~Figure~\ref{fig:system_act_md}c). This modification raises the degree of static indeterminacy to $n_\Rs = 2$ and leads to a slightly different motion with different evolutions of both elongation parameters. A significant difference, however, results from the fact the other bars cannot perform a pure rigid body motion anymore, but build up internal stresses throughout the motion. This is the main difference compared to the two previous motion designs and leads to a functional value of $J = 379$.

\subsection{Bridge structure with a traveling load}

\begin{figure}[b]\centering%\small
%\includesvg{fig_bridge}
\hspace*{-4mm}
\includegraphics[width=15.0cm]{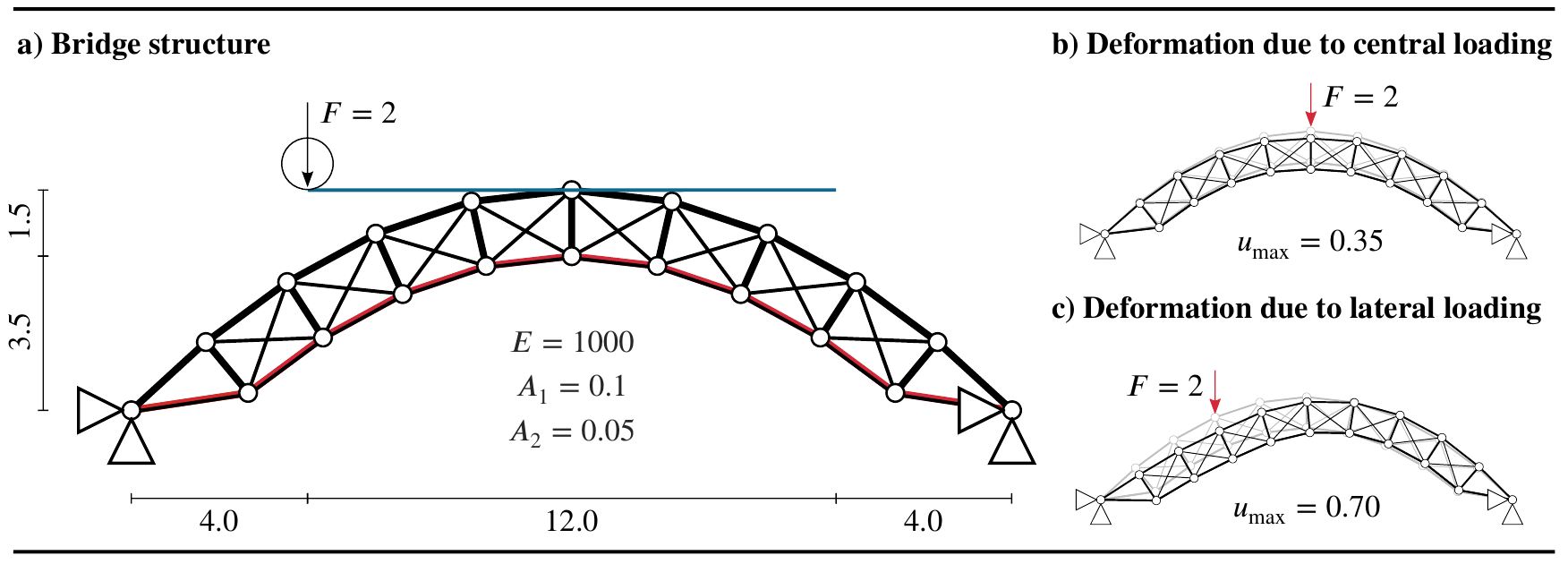}
\caption{Bridge structure and deformation due to static loading.}
\label{fig:bridge}
\end{figure}

To demonstrate the potential of motion design with actuator elements, a more complex example is presented in the following. A bridge is modeled as a truss structure, as shown in Figure~\ref{fig:bridge}a. Actuator elements are installed in the entire lower chord, thus resulting in a total of ten actuators. In this example, a parallel actuation mechanism is employed. This means that in addition to the actuator element, a regular, passive truss element is installed at the same place. As a result, this passive element is automatically stressed when the actuator changes its length. 
This is supposed to penalize the actuator extension in order to approximately represent the corresponding actuation costs.
%This leads to constraints during actuation, but represents an option to reflect the resistance of the actuator itself against a length change. 
The top chord and the vertical struts have a cross sectional area of $A_1 = 0.1$, while the diagonals as well as the actuator elements and the passive truss elements in the lower chord are built of elements with half the cross section area, i.e., $ A_2 =  0.05$. 

The bridge is exposed to a load $F$ traveling from left to right. The idea is that the point of load application remains at the same height throughout the entire travel, as indicated by the horizontal line in Figure~\ref{fig:bridge}a. In order to realize this, the structure has to adapt and undergo large deformations. The required motion should be solely enabled by the actuators in the lower chord.
%To obtain an estimation of the effect of the magnitude of the applied load, the passive structure without actuation is first loaded centrally and then laterally with a static point load. The deformed configurations can be seen in Figure~\ref{fig:bridge}b und Figure~\ref{fig:bridge}c, respectively. In both cases, significant nodal displacements of $u_\Rmax = 0.35$ and $u_\Rmax = 0.70$ are observed due to the static load. 

The initial configuration of the bridge is the symmetric, unloaded state. To find a suitable motion for the traveling load and to make the deformation process as efficient as possible, motion design is carried out in two stages. The two motion stages are not solved separately, but by one single motion design procedure. However, they differ in the controlled and prescribed displacement values, as described in the following.

\begin{figure}[b]\centering\small
%\includesvg{fig_bridge_sol}
\hspace*{-4mm}
\includegraphics[width=15.0cm]{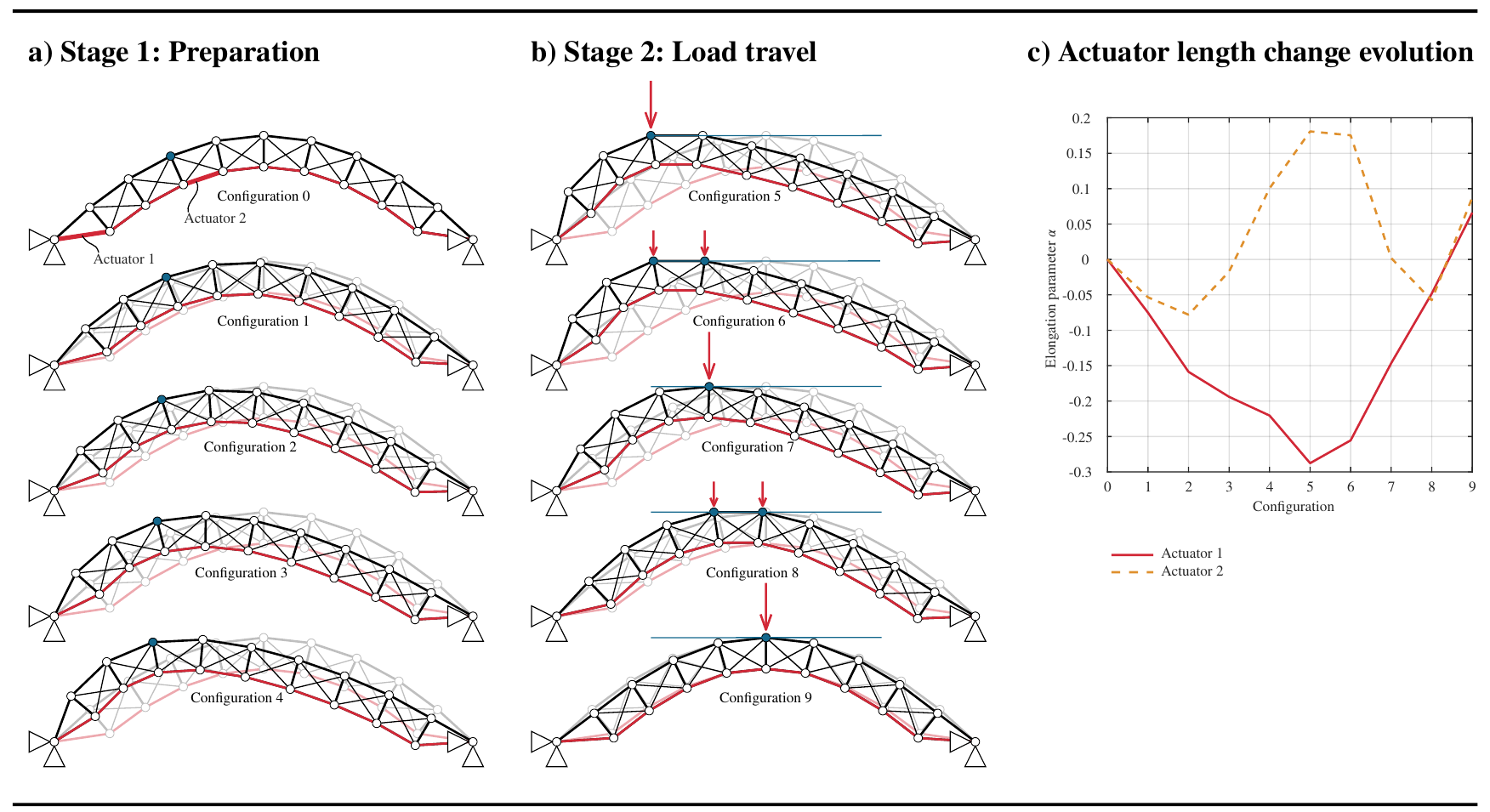}
\caption{Optimized motion of the bridge structure with traveling load.}
\label{fig:bridge_sol}
\end{figure}

Stage~1 of the motion represents the procedure of ``picking up'' of the point load on the left side. Therefore, the structure needs to deform sideways until the fourth node of the top chord (marked in blue) arrives at the starting point of the traveling load. For motion design, the vertical and horizontal target position of this node are prescribed 
and its horizontal displacement value is controlled throughout the motion of the first stage
(Configuration~0 to Configuration~5 in Figure~\ref{fig:bridge_sol}a).

In Stage~2, the load travels from the left side to the central node of the top chord. To guarantee that it always remains at the same vertical position, the vertical displacements of the nodes of the top chord are prescribed in a sequential way throughout the configuration: First, the vertical displacement of the fourth node is prescribed, then the vertical displacement of the fourth and fifth node are prescribed, then the vertical displacement of the fifth node is prescribed and so forth (cf. Figure~\ref{fig:bridge_sol}b, marked in blue). In the motion design problems described up to now, at least one target displacement value is prescribed for the motion and one degree of freedom is controlled throughout the entire motion. This is not the case in Stage~2, but the nodes with specified vertical displacement values change in every configuration. However, these specifications still lead to a stable motion design procedure with a unique result. Furthermore, the traveling load is applied to every node reaching this specific vertical position. If two nodes are placed at the same height, the load is split and distributed on both nodes. The application of these point loads is also included as a constraint on the motion as described in Section~\ref{sec:con_load_cases}. 

An alternative approach to design a motion is to find the geometry for every single configuration separately with the specific constraints, i.e., displacement values and loading, by an optimization that minimizes the internal energy. These optimized configurations can then be connected, thus generating an entire motion. These connected optimized geometries are then used as predictor motion for motion design. Figure~\ref{fig:bridge_sol}a and Figure~\ref{fig:bridge_sol}b show the optimized motion that results from the solution of the motion design problem including Stage~1 and Stage~2. Because of the applied constraints on the admissible point loads, the deformation is solely realized by the actuators. Furthermore, potential displacements due to the loading are compensated such that the point load can be kept at exactly the same height during the entire travel. The motions required to maneuver the load to the right side of the bridge is obtained by symmetry. Using the motion design method, the functional value reduces from $J=72.8$ in the predictor (connection of energy-minimal configurations) to $J=51.6$, i.e., by 29\%. This is due to the fact that the entire motion is considered in the objective function and not only separate configurations.

An output of the method are the targeted actuator length changes in the form of the elongation parameters, required to realize the optimal motion. Their evolution is exemplarily illustrated in Figure~\ref{fig:bridge_sol}c for the first actuator (Actuator~1) and the fourth actuator (Actuator~2) of the lower chord (marked in Configuration~0). It can be seen that the elongation parameters do not increase monotonously, but develop independently and can also decrease during the process. 
This confirms that the method of motion design allows to design efficient motions that are realized solely by actuator elements.

%%%%%%%%%%%%%%%%%%%%%%%%%%%%%%%%%%%%%%%%%%%%%%%%%%%%%%%%%%%%%%%%%

\section{Motion Stabilization}
\label{sec:con_detK}

\subsection{Detection of critical points}

In addition to the presented constraints regarding admissible actuation forces for the optimized motion, other constraints can be applied to motion design problems. One possibility is to enforce a stabilized motion, where no structural instabilities occur during the deformation. Instability phenomena, such as bifurcation or snap-through, are usually initiated at critical points, namely bifurcation points and limit points. Often, but not always, reaching a critical point is followed by entering an unstable equilibrium path. In quasi-static problems, the determinant of the tangent stiffness matrix is zero when a critical point is reached. This simple criterion is used in the following for identification of critical points.
% and becomes negative when this point is passed. If this happens, the deformed structure is in an unstable equilibrium state.

\subsection{Inequality constraint for the stiffness determinant}

The basic method for motion design again marks the starting point for constrained motion design, however, now with different constraint types. To avoid that the structure enters an unstable equilibrium path, an inequality constraint can be introduced within the framework indicated in eq.~(\ref{eq:func_ext_lc}). In particular, an inequality constraint can be applied onto the determinant of the tangent stiffness matrix $\BK_\RT$.
The constraints are summarized in a vector for all load configurations $\bar{k}$
\begin{align}
\Bg & = \begin{bmatrix} \det{\BK^2} & \det{\BK^3} & \ldots & \det{\BK^{\bar{k}}} & \ldots & \det{\BK^{\bar{n}_{\text{node}}}} \end{bmatrix}^\RT .
\end{align}
Thus, each entry $i$ of the vector $\Bg$, i.e., each constraint $g_i$, needs to fulfill the inequality\footnote{For a B-spline discretization, the nodes, where the constraints are enforced, are again the Greville collocation points on the deformation path itself.}
\begin{align}
g_i = \det{\BK_i} \ge 0 .
\end{align}
% \begin{align}
% g_i = -\det{\BK_i} \le 0 .
% \end{align}
% 
Applying inequality constraints is realized by activating a corresponding equality constrained, once the inequality constraint is violated. Hence, an active set strategy must be used. This is done by the Karush-Kuhn-Tucker conditions
\begin{align}
g_i & \ge 0 , &
\la_i & \ge 0 ,&
g_i \la_i = 0 
\end{align}
with the Lagrange multipliers $\la_i$. These conditions can also be expressed in a semi-smooth way in the complementarity function for every constraint $i$
\begin{align}
C_i = \la_i - \max{(0,\la_i-c g_i)} = 0 .
\end{align}
The complementarity parameter $c$ has to be positive and stabilizes the system of equations.

Like in Section~\ref{sec:con_load_cases}, the derivatives of the constraints have to be computed to set up the extended system of equations~\ref{eq:system_eq_ext}. Clearly, this is a non-trivial task, but it can also be accomplished numerically. Also here, the complex step differentiation from \cite{martins_complex-step_2003} is used to get the exact derivatives. However, only the first derivative can be calculated exactly in this way. 
%\textcolor{red}{\sout{To avoid the calculation of the second derivative, it can simply be suspended. This leads to a modified system of equations without the Hessian matrix $\BH$}} 
A possibility to numerically compute second derivates is to make use of hyper-dual numbers, as proposed by Fike and Alonso \cite{fike_development_2011}. Even though these numbers enable the computation of exact second derivates, they also suffer from the drawback that they are not included in the standard C++ library and are therefore not accessible for a variety of finite element research software. As an alternative, the calculation of the Hessian and the associated second derivatives of the determinant can simple be omitted. This leads to a not consistently linearized problem and the Newton-Raphson method is transformed into a modified Newton-Raphson solution scheme. It follows for the resulting system of equations
\begin{align}
\begin{bmatrix}
\BK_{\Rm\Rd} + \cancel{\la_i \BH_i} & \BG^\RT \\ \BG & \B0
\end{bmatrix}
\begin{bmatrix}
 \DE \bar{\BD} \\ \DE \Bla
\end{bmatrix} = - \begin{bmatrix}
\BR_{\Rm\Rd} + \Bla^\RT \BG \\ \Bg
 \end{bmatrix} .
 \label{eq:system_eq_ext_detK}
\end{align}
%
%\textcolor{red}{\sout{Thus, the Newton-Raphson method is transformed into a modified Newton-Raphson solution scheme, in which the problem is not consistently linearized.}} 
The incomplete linearization may affect the convergence behavior of the non-linear problem. It is expected to have an inferior, no longer quadratic convergence behavior, compared with the complete system of equations including the Hessian matrix. Nevertheless, the convergence behavior is expected to remain acceptable, since $\BK_{\Rm\Rd}$ is still updated in each iteration. Due to the correct residual, the solution still converges to the correct minimal solution, but usually with an increased number of iterations. %\textcolor{red}{\sout{Alternatively, another possibility to set up the system of equations with consistent linearization is to calculate the second derivative exactly with hyper-dual numbers as proposed by \cite{fike_development_2011}.}}

To demonstrate the stabilization of motions, two simple two-bar structures, incorporating snap-through phenomena as well as bifurcation points are presented in the following.

\subsection{Snap-through problem in a shallow two-bar truss}

\begin{figure}[b]
\centering\small
%\includesvg{fig_system_detK}
\hspace*{-4mm}
\includegraphics[width=15.0cm]{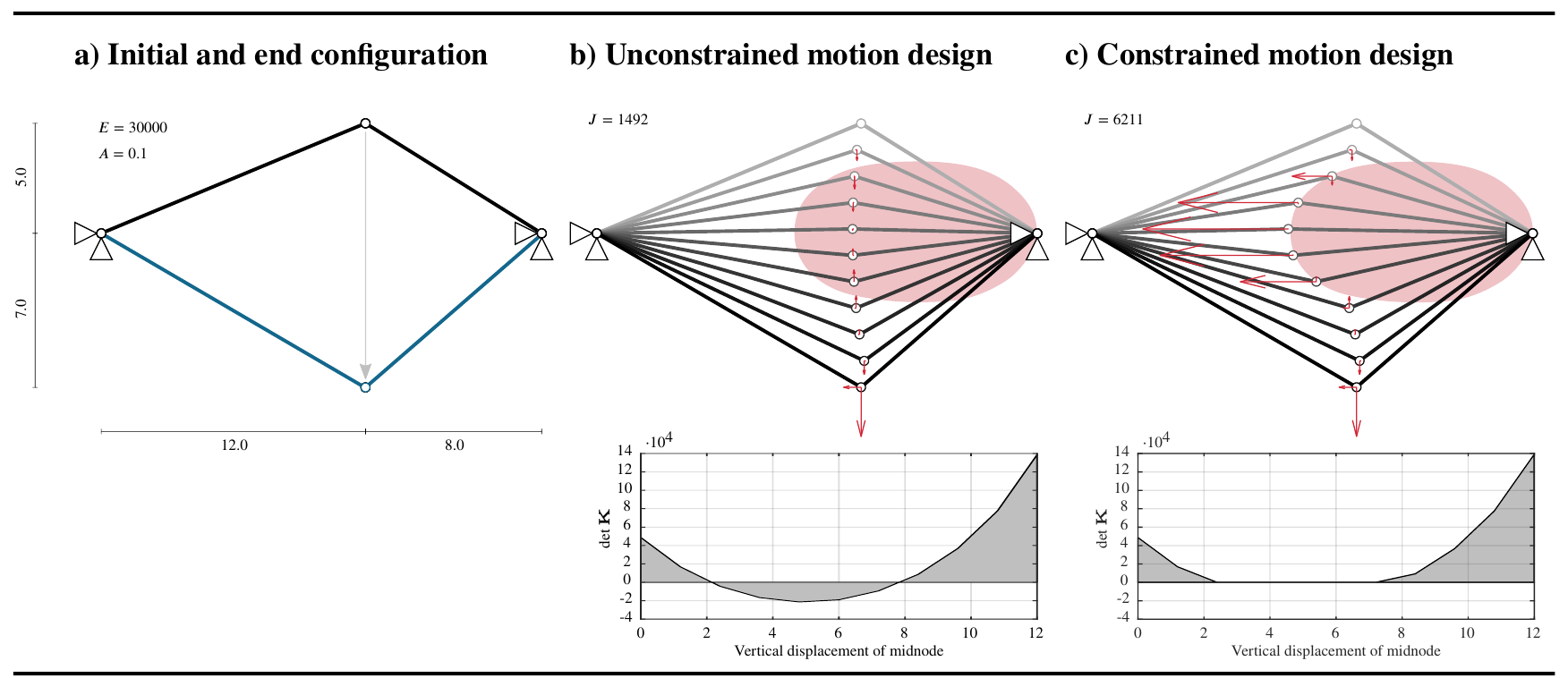}
\caption{Unsymmetric two bar truss with an unconstrained and constrained stabilized motion design.}
\label{fig:system_detK}
\end{figure}

The method of stabilized motion design is first investigated on the example of an unsymmetric two-bar truss, as shown in Figure~\ref{fig:system_detK}a.
%The asymmetry is chosen to avoid numerical instabilities in the calculation attributable to multiple possible solution curves.
Here, no constraints are enforced regarding the actuation forces. Thus, a horizontal and a vertical force can be applied onto the free node. This corresponds to an unconstrained motion design, the result of which is shown in Figure~\ref{fig:system_detK}b for a path discretization with ten elements and the vertical displacement of the central node being controlled. The sign of $\det \BK_\RT$ is indicated in the background of the illustrated motion, where the red shaded zone marks a region in which $\det \BK_\RT < 0$. In the unconstrained motion design solution, the node traverses this unstable region. This is also visible in the plot of the determinant of the stiffness matrix versus the displacement $D_2$ in Figure~\ref{fig:system_detK}b. Between vertical displacement values of approximately $D_2\approx 2$ and $D_2\approx 8$, the determinant takes a negative value. For the given load-controlled process, this would practically mean that an uncontrolled dynamic snap-through occurs instead of the desired (designed) motion. 

Applying constrained motion design, a stabilized motion is obtained (see Figure~\ref{fig:system_detK}c). The trajectory of the central node only touches the zone where the determinant would become negative. In practical applications, one would probably avoid exactly reaching this critical region, but keep a safe distance by enforcing a (small) positive value $\det \BK_\RT > 0$ in the inequality constraint.

It can also be seen that the magnitude of the forces, especially the horizontal point load at the midnode, changes in order to follow the alternative deformation path. The motion design functional itself, i.e., the minimum of the cost of deformation, stays the same, but its value increases drastically from $J=1492$ to $J=6211$ due to the necessary detour. However, a stabilization and a positive determinant are ensured throughout the entire motion.
%This is achieved by the initial stress stiffness or geometric stiffness. In a typical snap-through process, this part of the stiffness becomes negative and compensates the elastic and initial displacement stiffness contributions. In the stable motion, the geometrical stiffness is balanced by the combined tension and pressure normal force of the two bars. Like this, its negative value does not dominate the two other parts of the whole stiffness and the determinant does not decrease to a value lower than zero.

\subsection{Stabilization of a two-bar truss with bifurcation}

%When the midnode of a shallow two-bar truss is moved downwards, the limit point is the starting point for the snap-through process. In a high two-bar truss, an additional bifurcation point occurs before the limit point is reached. This happens when the height to the total span ratio exceeds the value of $\sqrt{2}/2$. Therefore a higher two-bar structure is modeled as shown in Figure~\ref{fig:system_detK_bif}a. 

\begin{figure}[t]
\centering\small
%\includesvg{fig_system_detK_bif}
\hspace*{-4mm}
\includegraphics[width=15.0cm]{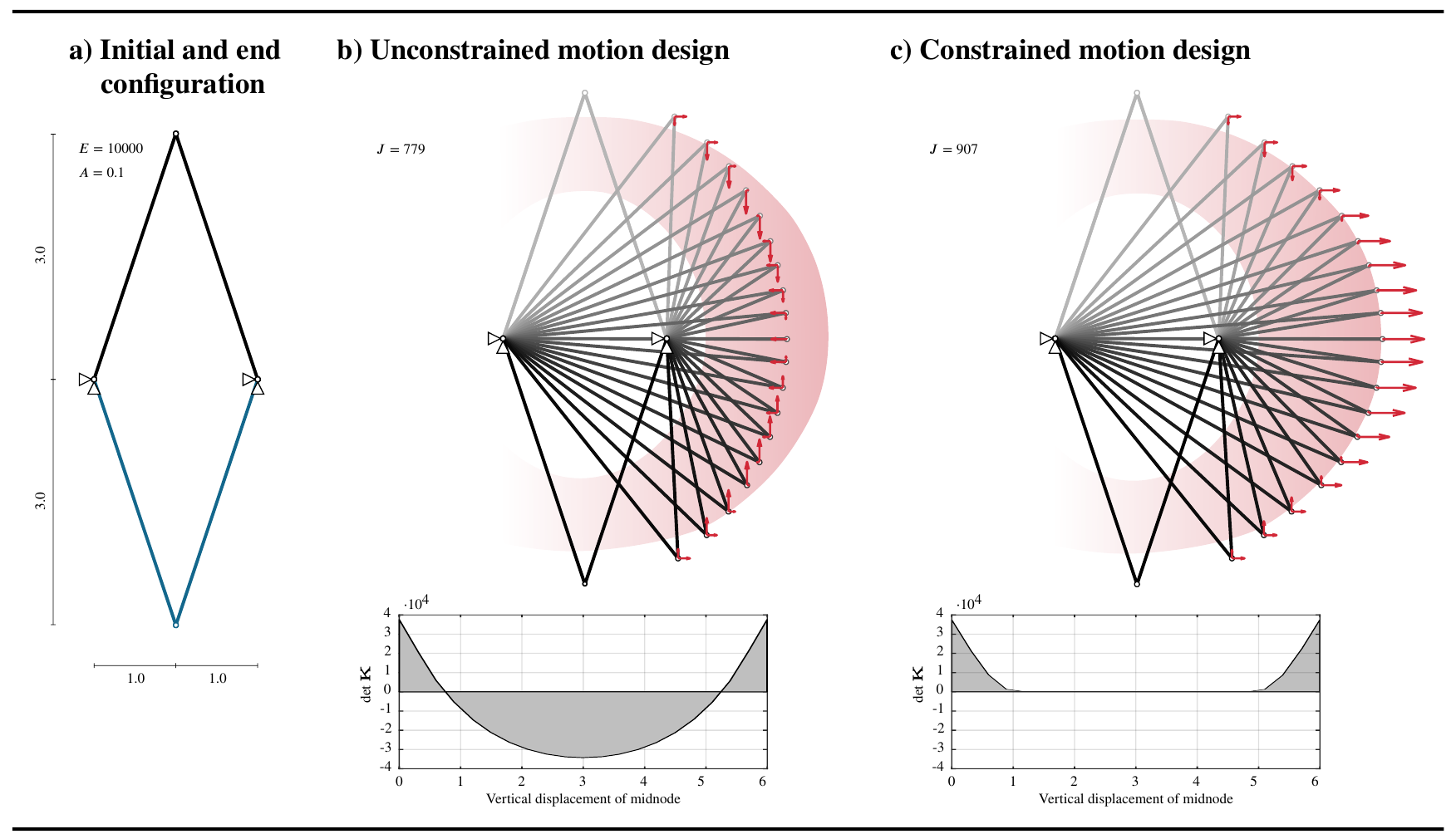}
\caption{High two-bar truss with a bifurcation point and an unstabilized and stabilized motion.}
\label{fig:system_detK_bif}
\end{figure}

In order to also investigate the stabilization in bifurcation problems, a high two-bar truss is studied next (see Figure~\ref{fig:system_detK_bif}a. When the free node of a \emph{shallow} two-bar truss is moved downwards, the first critical point reached is a limit point, initiating a snap-through process. However, in this case, an additional bifurcation point occurs before the limit point is reached. This happens when the height to span ratio exceeds a value of $\sqrt{2}/2$. The objective of motion design is again to reach a ``mirrored'' configuration. 

This problem is already described in~\cite{sachse_motion_2019}, where different solutions and equilibrium paths are compared. Figure~\ref{fig:system_detK_bif}b shows the solution from unconstrained motion design (with twenty linear path elements and the vertical displacement controlled). Again, the red zone indicates a negative sign of the determinant of the tangent stiffness matrix $\det{\BK_\RT}$.
%If the central node is located in this zone, the determinant becomes negative and the deformed structure is characterized by an unstable equilibrium state with the corresponding external forces.
%However, another white zone ($\det{\BK_\RT \ge 0}$) is located within the unstable region. This is a result of the limit point that is only reached after the bifurcation point has been passed. The two resulting negative eigenvalues of the tangent stiffness matrix, again indicating an unstable deformation state, compensate each other. Therefore, the determinant becomes positive again, although no stable state has been reached.
Again, the unstable region is traversed by the deformation path, which can also be seen in the progress of the stiffness determinant in the corresponding diagram underneath. Thus, this result would be practically infeasible.

Applying the inequality constraint $\det{\BK_\RT \ge 0}$ results in the motion that can be seen in Figure~\ref{fig:system_detK_bif}c, where the red zone is again only touched by the free node, but not traversed. The constraint of enforcing a positive stiffness determinant can therefore also be successfully applied to problems where bifurcation phenomena can occur.

\subsection{Stabilization of a motion with restricted actuation}
\label{sec:con_lc_detK}

So far, the two constraint types of enabling an optimized motion by a restricted set of actuation forces (equality constraints) and of stabilizing the motion by enforcing a positive stiffness determinant (inequality constraints) were treated separately. However, they can also be combined. The vector $\Bg$ that gathers all constraints then contains entries with the internal forces $\Bg_F$ as well as entries with the determinant of the stiffness matrix $\Bg_{\det}$, 
\begin{align}
\Bg = \begin{bmatrix} \Bg_F \\ \Bg_{\det} \end{bmatrix} \, .
\end{align}
As a consequence, the resulting system of equations is further expanded. This is shown in the following with the system from Figure~\ref{fig:system_lc}. The deformation path is discretized by 16 linear elements (only eight are illustrated in Figure~\ref{fig:system_lc_sol}) and the vertical displacement is controlled. 
It is already known at this point that unconstrained motion design yields the functional value $J=600$, as shown in Figure~\ref{fig:system_lc_sol}a. This represents the reference value for the following modifications. 

\begin{figure}[t]
\centering\small
%\includesvg{fig_system_lc_detK}
\hspace*{-4mm}
\includegraphics[width=15.0cm]{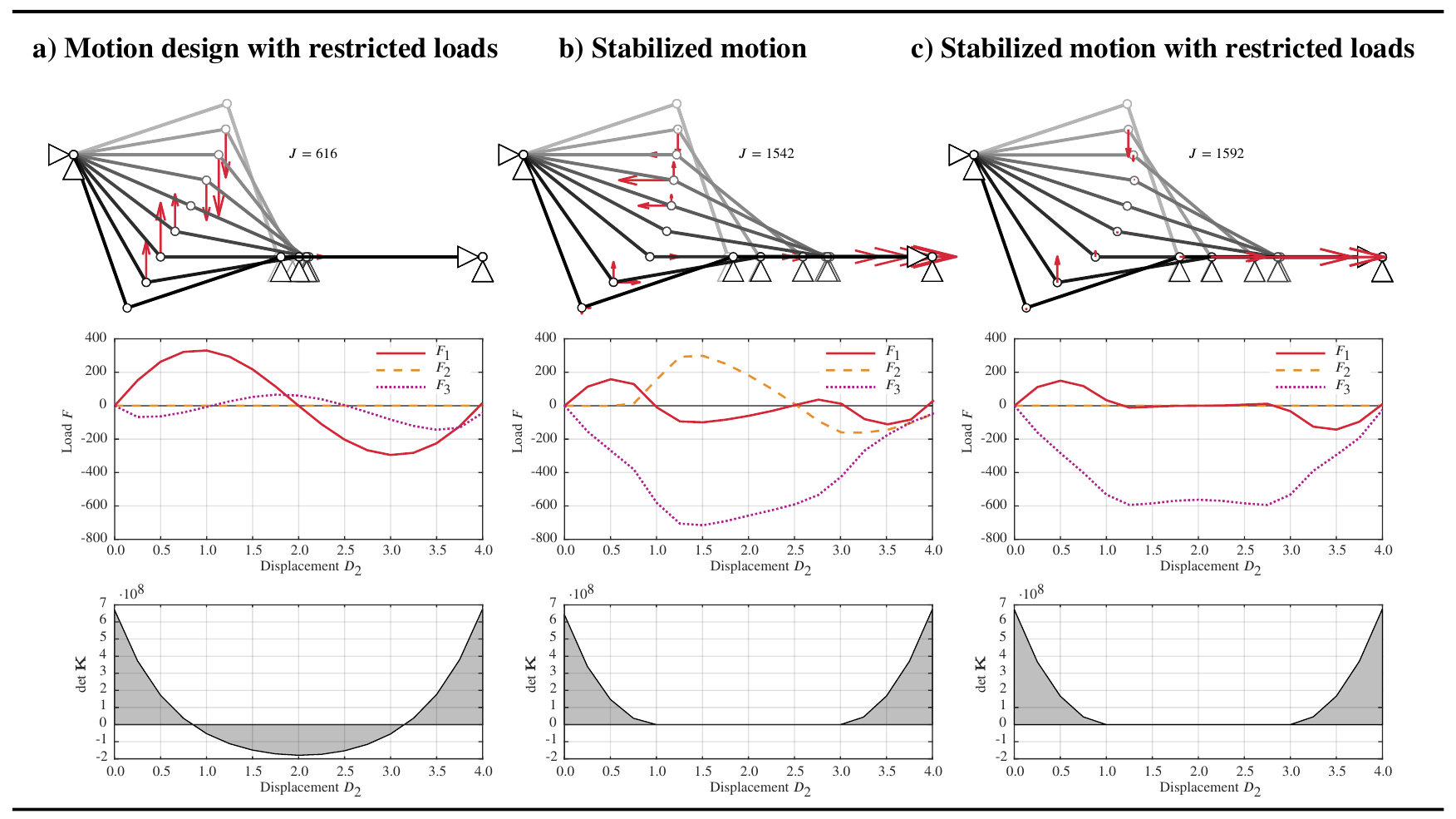}
\caption{Constrained motion design on an exemplary system with different types of constraints as well as their combination.}
\label{fig:system_lc_detK}
\end{figure}

First, the horizontal force at the free node is suspended and a constrained motion design is carried out. This leads to the motion in Figure~\ref{fig:system_lc_detK}a. Due to the constraint, the value of the functional increases to $J=616$. Secondly, all possible forces may again be applied, but a positive determinant of the stiffness matrix is enforced to find a stabilized motion. Again, a different motion is found, shown in Figure~\ref{fig:system_lc_detK}b, and the functional value increases significantly to $J=1542$. In this structure, the horizontal bar is activated much more in the form of a high compressive force to reduce the snap-through effect of the two other bars.
%This result incorporates the same considerations about the initial stress stiffness as in the previous example of a shallow two-bar truss.
Finally, both constraints can be combined. The result is a stable motion that always keeps a positive determinant of the stiffness matrix, but is also realizable with only two point forces, as illustrated in Figure~\ref{fig:system_lc_detK}c.  

%%%%%%%%%%%%%%%%%%%%%%%%%%%%%%%%%%%%%%%%%%%%%%%%%%%%%%%%%%%%%

\section{Conclusions}
\label{sec:conclusions}

In this paper, the work presented in Sachse and Bischoff \cite{sachse_motion_2019} is continued and extended to a more practical approach. While in the previous study, only unconstrained motions were investigated, this work presents the introduction of constraints into the motion design process. This allows finding an optimal motion between two prescribed geometries that is either realized by only a specific number of load cases, actuator elements, or is stabilized by enforcing a positive structure's stiffness determinant.

After the basics of the unconstrained motion design method were summarized, first, the introduction of load cases to motion design was presented. This was introduced by enforcing a zero internal force at every unloaded degree of freedom and can exemplarily be conducted with the Lagrange multiplier method. Here, the second derivatives of the constraints need to be calculated to build the resulting extended system of equations. The first derivative is represented by the stiffness matrix, whereas the second derivative, the derivative of the stiffness, can be obtained by a directional derivative. Due to the particular (symmetric) character of the stiffness matrix, this directional derivative allows for using only one more evaluation of the function when applying numerical differentiation, increasing the numerical efficiency of the calculation. Furthermore, additional considerations have to be made for the definition of the desired end geometry. This approach is further demonstrated in some numerical experiments with different types of finite elements. They showed that the functional value of constrained optimized motions increases compared to unconstrained optimized motions. However, the constraints can be satisfied.

The length change of incorporated actuator elements in the structure can also be regarded as a load case. Therefore, a new actuator element formulation with an additional parameter, the targeted actuator length change $\al$, was developed. This element formulation allows a straightforward implementation into the presented motion design method because it is simply solved for this additional parameter as well. By application of the actuator element, kinematic motion of the remaining passive structure and the corresponding actuator length changes can be determined. In addition, the potential of this method with actuator elements is shown on the example of a bridge structure with a traveling load. 

Next, the constraint of a stabilized motion is introduced. In this case, the fact that the determinant of the stiffness matrix of the structure becomes negative when passing a critical point is utilized. The inequality condition of always keeping a positive determinant is also enforced with the Lagrange multiplier method while following an active set strategy. However, the required derivatives are difficult to obtain. This is why, again, numerical differentiation is used. To improve efficiency, only the first derivatives of the constraints are calculated in this way leading to a modified Newton scheme. Due to the type of constraint, where a determinant needs to be calculated and differentiated, the solution process might be inefficient in terms of the calculation time for large structures. Furthermore, the solution of the resulting extended system of equations might exhibit numerical instabilities as the entries are characterized by a significant difference in magnitude, which is caused by the determinant's value. However, in this study, only the principle of the introduction of such a constraint is shown and its functionality proven with simple structures. When the presented approach is applied in the presented numerical examples, unstable deformation states are avoided in snap-through problems as well as bifurcation problems. The described numerical problems still need to be tackled in further research if following this idea.

Finally, both type of constraints, enabling the designed motion only by a prescribed number of loads as well as the motion stabilization, is combined and presented in an exemplary structure. Similarly, other inequality constraints on stress measures are thinkable. This way, it can be ensured that they do not exceed any strength values or limits regarding local instability behavior, i.e., buckling of single truss elements, during the entire deformation process. For such constraints, usually only existing quantities in a nonlinear finite element code, such as the first and second derivatives of the stress tensor, are required and analytical sensitivities without the need for numerical differentiation can be obtained.

The basic concept and structure of unconstrained and constrained motion design is presented with an exemplary functional, the integrated internal energy over the deformation path, and exemplary constraints in the previous publication by the authors \cite{sachse_motion_2019} as well as in this paper. The herein described constraints should represent an approach for technical implementation and application of the motion design method for real structures. However, also other functionals and constraints may be introduced to adapt the method for the specific requirements of the underlying task.

Besides, the type and number of loads were always defined prior to motion design. One possibility for further research is to include an optimization procedure also to find the optimal loads to trace the optimized deformation path by motion design. This resembles actor placement procedures or, in this case, load case placement procedures, that already exist as described in the introductory section~\ref{sec:intro}. Furthermore, the focus was laid on the application of point forces up to now, but also other types of actuation are thinkable such as discrete actuator elements or piezoelectric structural parts and may be implemented.

%%%%%%%%%%%%%%%%%%%%%%%%%%%%%%%%%%%%%%%%%%%%%%%%%%%%%%%%%%%%%

\section*{Acknowledgements}

This research work by Renate Sachse was funded and supported by the State Ministry of Baden-Wuerttemberg for Sciences, Research and Arts.
Furthermore, the contribution of Florian Geiger was conducted in the framework of the Collaborative Research Centre 1244 “Adaptive Skins and Structures for the Built Environment of Tomorrow”/project B01 funded by the Deutsche Forschungsgemeinschaft (DFG, German Research Foundation) - project number 279064222. The authors are grateful for the generous support.

%%%%%%%%%%%%%%%%%%%%%%%%%%%%%%%%%%%%%%%%%%%%%%%%%%%%%%%%%%%%%

\bibliographystyle{plain}
\bibliography{literature}

\begin{thebibliography}{10}

\bibitem{abdullah_placement_2001}
Makola~M. Abdullah, Andy Richardson, and Jameel Hanif.
\newblock Placement of sensors/actuators on civil structures using genetic
  algorithms.
\newblock {\em Earthquake Engineering \& Structural Dynamics},
  30(8):1167--1184, 2001.

\bibitem{ajaj_morphing_2016}
Rafic~M. Ajaj, Christopher~S. Beaverstock, and Michael~I. Friswell.
\newblock Morphing aircraft: {The} need for a new design philosophy.
\newblock {\em Aerospace Science and Technology}, 49:154--166, February 2016.

\bibitem{liebe_shape-adaptive_2006}
L.F. Campanile.
\newblock Shape-adaptive wings—the unfulfilled dream of flight.
\newblock In R.~Liebe, editor, {\em {WIT} {Transactions} on {State} of the
  {Art} in {Science} and {Engineering}}, volume~2, pages 400--419. WIT Press, 1
  edition, November 2006.

\bibitem{campanile_modal_2008}
L.F. Campanile.
\newblock Modal {Synthesis} of {Flexible} {Mechanisms} for {Airfoil} {Shape}
  {Control}.
\newblock {\em Journal of Intelligent Material Systems and Structures},
  19(7):779--789, July 2008.

\bibitem{fike_development_2011}
Jeffrey Fike and Juan Alonso.
\newblock The {Development} of {Hyper}-{Dual} {Numbers} for {Exact}
  {Second}-{Derivative} {Calculations}.
\newblock In {\em 49th {AIAA} {Aerospace} {Sciences} {Meeting} including the
  {New} {Horizons} {Forum} and {Aerospace} {Exposition}}, Orlando, Florida,
  January 2011. American Institute of Aeronautics and Astronautics.

\bibitem{goppert_spoked_2007}
Knut Göppert and Michael Stein.
\newblock A {Spoked} {Wheel} {Structure} for the {World}’s largest
  {Convertible} {Roof} – {The} {New} {Commerzbank} {Arena} in {Frankfurt},
  {Germany}.
\newblock {\em Structural Engineering International}, 17(4):282--287, 2007.

\bibitem{graells_rovira_control_2009}
Albert Graells~Rovira and Josep~M. Mirats~Tur.
\newblock Control and simulation of a tensegrity-based mobile robot.
\newblock {\em Robotics and Autonomous Systems}, 57(5):526--535, 2009.

\bibitem{gupta_optimization_2010}
Vivek Gupta, Manu Sharma, and Nagesh Thakur.
\newblock Optimization {Criteria} for {Optimal} {Placement} of {Piezoelectric}
  {Sensors} and {Actuators} on a {Smart} {Structure}: {A} {Technical} {Review}.
\newblock {\em Journal of Intelligent Material Systems and Structures},
  21(12):1227--1243, August 2010.

\bibitem{housner_structural_1997}
G.~W. Housner, L.~A. Bergman, T.~K. Caughey, A.~G. Chassiakos, R.~O. Claus,
  S.~F. Masri, R.~E. Skelton, T.~T. Soong, B.~F. Spencer, and J.~T.~P. Yao.
\newblock Structural {Control}: {Past}, {Present}, and {Future}.
\newblock {\em Journal of Engineering Mechanics}, 123(9):897--971, September
  1997.

\bibitem{ibrahimbegovic_optimal_2004}
A.~Ibrahimbegovic, C.~Knopf‐Lenoir, A.~Kučerová, and P.~Villon.
\newblock Optimal design and optimal control of structures undergoing finite
  rotations and elastic deformations.
\newblock {\em International Journal for Numerical Methods in Engineering},
  61(14):2428--2460, 2004.

\bibitem{balaguer_development_2008}
Fumihiro Inoue.
\newblock Development of {Adaptive} {Construction} {Structure} by {Variable}
  {Geometry} {Truss}.
\newblock In Carlos Balaguer and Mohamed Abderrahim, editors, {\em Robotics and
  {Automation} in {Construction}}. InTech, October 2008.

\bibitem{irschik_review_2002}
H.~Irschik.
\newblock A review on static and dynamic shape control of structures by
  piezoelectric actuation.
\newblock {\em Engineering Structures}, 24(1):5--11, January 2002.

\bibitem{korkmaz_review_2011}
Sinan Korkmaz.
\newblock A review of active structural control: challenges for engineering
  informatics.
\newblock {\em Computers \& Structures}, 89(23):2113--2132, 2011.

\bibitem{kota_design_2001}
Sridhar Kota, Jinyong Joo, Zhe Li, Steven~M. Rodgers, and Jeff Sniegowski.
\newblock Design of {Compliant} {Mechanisms}: {Applications} to {MEMS}.
\newblock {\em Analog Integrated Circuits and Signal Processing}, 29(1):7--15,
  October 2001.

\bibitem{lienhard_flectofin:_2011}
J~Lienhard, S~Schleicher, S~Poppinga, T~Masselter, M~Milwich, T~Speck, and
  J~Knippers.
\newblock Flectofin: a hingeless flapping mechanism inspired by nature.
\newblock {\em Bioinspiration \& Biomimetics}, 6(4):045001, 2011.

\bibitem{lu_effective_2005}
Kerr-Jia Lu and Sridhar Kota.
\newblock An {Effective} {Method} of {Synthesizing} {Compliant} {Adaptive}
  {Structures} using {Load} {Path} {Representation}.
\newblock {\em Journal of Intelligent Material Systems and Structures},
  16(4):307--317, April 2005.

\bibitem{martins_complex-step_2003}
Joaquim R. R.~A. Martins, Peter Sturdza, and Juan~J. Alonso.
\newblock The {Complex}-step {Derivative} {Approximation}.
\newblock {\em ACM Trans. Math. Softw.}, 29(3):245--262, September 2003.

\bibitem{masching_parameter_2016}
Helmut Masching and Kai-Uwe Bletzinger.
\newblock Parameter free structural optimization applied to the shape
  optimization of smart structures.
\newblock {\em Finite Elements in Analysis and Design}, 111:33--45, April 2016.

\bibitem{masic_path_2005}
Milenko Masic and Robert~E. Skelton.
\newblock Path {Planning} and {Open}-{Loop} {Shape} {Control} of {Modular}
  {Tensegrity} {Structures}.
\newblock {\em Journal of Guidance, Control, and Dynamics}, 28(3):421--430, May
  2005.

\bibitem{maute_integrated_2006}
Kurt~K. Maute and Gregory~W. Reich.
\newblock Integrated {Multidisciplinary} {Topology} {Optimization} {Approach}
  to {Adaptive} {Wing} {Design}.
\newblock {\em Journal of Aircraft}, 43(1):253--263, January 2006.

\bibitem{pagitz_shape-changing_2013}
M~Pagitz and J~Bold.
\newblock Shape-changing shell-like structures.
\newblock {\em Bioinspiration \& Biomimetics}, 8(1):016010, February 2013.

\bibitem{reksowardojo_actuator_2018}
Arka~P. Reksowardojo, Gennaro Senatore, and Ian F.~C. Smith.
\newblock Actuator {Layout} {Optimization} for {Adaptive} {Structures}
  {Performing} {Large} {Shape} {Changes}.
\newblock In Ian F.~C. Smith and Bernd Domer, editors, {\em Advanced
  {Computing} {Strategies} for {Engineering}}, Lecture {Notes} in {Computer}
  {Science}, pages 111--129. Springer International Publishing, 2018.

\bibitem{rus_design_2015}
Daniela Rus and Michael~T. Tolley.
\newblock Design, fabrication and control of soft robots.
\newblock {\em Nature}, 521(7553):467--475, May 2015.

\bibitem{sachse_motion_2019}
Renate Sachse and Manfred Bischoff.
\newblock A variational formulation for motion design of adaptive compliant
  structures.
\newblock {\em International Journal for Numerical Methods in Engineering},
  Available online at: https://onlinelibrary.wiley.com/doi/abs/10.1002/nme.6570
  (Accessed December 5,2020), 2020.

\bibitem{senatore_synthesis_2019}
Gennaro Senatore, Philippe Duffour, and Peter Winslow.
\newblock Synthesis of minimum energy adaptive structures.
\newblock {\em Structural and Multidisciplinary Optimization}, March 2019.

\bibitem{sigmund_design_1997}
Ole Sigmund.
\newblock On the {Design} of {Compliant} {Mechanisms} {Using} {Topology}
  {Optimization}.
\newblock {\em Mechanics of Structures and Machines}, 25(4):493--524, January
  1997.

\bibitem{sobek_adaptive_2001}
Werner Sobek and Patrick Teuffel.
\newblock Adaptive systems in architecture and structural engineering.
\newblock In {\em Smart {Structures} and {Materials} 2001: {Smart} {Systems}
  for {Bridges}, {Structures}, and {Highways}}, volume 4330, pages 36--46.
  International Society for Optics and Photonics, 2001.

\bibitem{spencer_b._f._state_2003}
{Spencer B. F.} and {Nagarajaiah S.}
\newblock State of the {Art} of {Structural} {Control}.
\newblock {\em Journal of Structural Engineering}, 129(7):845--856, July 2003.

\bibitem{sychterz_deployment_2018}
Ann~C. Sychterz and Ian F.~C. Smith.
\newblock Deployment and {Shape} {Change} of a {Tensegrity} {Structure} {Using}
  {Path}-{Planning} and {Feedback} {Control}.
\newblock {\em Frontiers in Built Environment}, 4, 2018.

\bibitem{vasista_realization_2012}
Srinivas Vasista, Liyong Tong, and K.~C. Wong.
\newblock Realization of {Morphing} {Wings}: {A} {Multidisciplinary}
  {Challenge}.
\newblock {\em Journal of Aircraft}, 49(1):11--28, January 2012.

\bibitem{veuve_adaptive_2017}
Nicolas Veuve, Ann~C. Sychterz, and Ian~F.C. Smith.
\newblock Adaptive control of a deployable tensegrity structure.
\newblock {\em Engineering Structures}, 152:14--23, December 2017.

\bibitem{weisshaar_morphing_2013}
Terrence~A. Weisshaar.
\newblock Morphing {Aircraft} {Systems}: {Historical} {Perspectives} and
  {Future} {Challenges}.
\newblock {\em Journal of Aircraft}, 50(2):337--353, March 2013.

\bibitem{wijdeven_shape_2005}
J.~van~de Wijdeven and B.~de Jager.
\newblock Shape change of tensegrity structures: design and control.
\newblock In {\em Proceedings of the 2005, {American} {Control} {Conference},
  2005.}, pages 2522--2527 vol. 4, June 2005.

\end{thebibliography}

%%%%%%%%%%%%%%%%%%%%%%%%%%%%%%%%%%%%%%%%%%%%%%%%%%%%%%%%%%%%%

%\appendix

%%%%%%%%%%%%%%%%%%%%%%%%%%%%%%%%%%%%%%%%%%%%%%%%%%%%%%%%%%%%%

\end{document}